\documentclass[floatfix,aps,twocolumn,showpacs,superscriptaddress,pra,longbibliography,10pt]{revtex4-2}
\usepackage{amsmath,amsfonts,amssymb,graphicx}
\usepackage{subfigure}
\usepackage{epstopdf}
\usepackage{bm}
\usepackage{hyperref}
\usepackage{comment}
\usepackage[utf8]{inputenc}
\usepackage[T1]{fontenc}
\usepackage[english]{babel}
\IfFileExists{newtxtext.sty}
    {\usepackage{newtxtext,newtxmath}}
    {\IfFileExists{stix.sty}
       {\usepackage{stix}}
       {\IfFileExists{mathptmx.sty}
       {\usepackage{mathptmx}}{} } }

\newcommand{\sgn}{\mathrm{sgn}}

\begin{document}

\title{Quantum Monte Carlo and perturbative study of two-dimensional Bose-Fermi mixtures}

\author{Jacopo D'Alberto}
\affiliation{Dipartimento di Fisica ``Aldo Pontremoli'', Universit\`a degli Studi di Milano, via Celoria 16, I-20133 Milano, Italy}
\author{Lorenzo Cardarelli}
\affiliation{PASQAL, 7 rue L\'eonard de Vinci, 91300 Massy, France}
\author{Davide Emilio Galli}
\affiliation{Dipartimento di Fisica ``Aldo Pontremoli'', Universit\`a degli Studi di Milano, via Celoria 16, I-20133 Milano, Italy}
\author{Gianluca Bertaina}
\email{g.bertaina@inrim.it}
\affiliation{Istituto Nazionale di Ricerca Metrologica, Strada delle Cacce 91, I-10135 Torino, Italy} 
\author{Pierbiagio Pieri}
\email{pierbiagio.pieri@unibo.it}
\affiliation{Dipartimento di Fisica e Astronomia ``Augusto Righi", Università di Bologna, Via Irnerio 46, I-40126, Bologna, Italy}
\affiliation{INFN, Sezione di Bologna,  Viale Berti Pichat 6/2, I-40127, Bologna, Italy}

\begin{abstract}
We derive analytically the leading beyond-mean field contributions to the zero-temperature equation of state and to the fermionic quasi-particle residue  and effective mass of a dilute Bose-Fermi mixture in two dimensions. In the repulsive case, we perform quantum Monte Carlo simulations  for two representative bosonic concentrations and equal masses, extending a method for correcting finite-size effects in fermionic gases to Bose-Fermi mixtures. We find good agreement between analytic expressions and numerical results for weak interactions, while significant discrepancies appear in the regime close to mechanical instability, above which we provide evidence of phase separation of the bosonic component. 
\end{abstract}

\maketitle

\section{Introduction}
Quantum mixtures have attracted considerable interest in recent years. Thanks to their versatility and tunability they turned out to be an ideal platform to test and develop new physics \cite{Bloch_Rev_2008}. In particular, Bose-Bose (BB) mixtures and Fermi-Fermi (FF) mixtures have been studied in depth both from a theoretical and experimental point of view, resulting in a better understanding of important quantum phenomena such as fermionic superfluidity and the BCS-BEC crossover in FF mixtures \cite{Giorgini_Rev_2008,Strinati-2018} or self-bound quantum droplets in BB mixtures \cite{Semeghini-2018,Cabrera-2018}. 

Bose-Fermi (BF) mixtures have also been investigated, although somewhat less extensively in the literature.
Initial theoretical studies of BF mixtures considered the problem of instability 
(by collapse or phase separation) using mean-field and perturbative approaches, emphasizing the need for a sufficiently high repulsive interaction between bosons \cite{Viverit_Stability_2000,Roth_Instability_2002,Roth_Stability_2002}. The occurrence of instabilities was also confirmed experimentally by the first realization of a non-resonant BF mixture \cite{Modugno_Instability_2002}. Later, with the development and refinement of more sophisticated techniques such as optical lattices and Feshbach resonances \cite{Lewenstein_OpticalLattices_2007,Chin_Feshbach_2010}, the focus shifted to strongly interacting Bose-Fermi systems. Indeed, the possibility to tune the interaction between bosons and fermions by varying an external magnetic field has ensured different possible implementation scenarios.
    
The case of a Bose-Fermi mixture with a BF interaction modulated by a Feshbach resonance has attracted particular interest in the literature. Initial works studying this system in the presence of a resonance focused on lattice models \cite{Kagan_Lattice_2004,Gunter_Lattice_2006,Barillier_Lattice_2008}. Subsequent works moved instead to the continuous case \cite{Watanabe_ContinuousBF_2008}. In addition to the instability problem \cite{Yu_Stability_2011,Ludwig_Stability_2011}, the competition between BF pairing and boson condensation in a Bose-Fermi mixture across a broad Feshbach resonance was studied in \cite{Fratini_Pairing_2010,Bertaina_BF3D_2013, Guidini_CondensedphaseBoseFermi_2015}. In particular, it was found that for weak attraction, at sufficiently low temperature, the bosons condense while the fermions behave like a Fermi liquid. Instead, for sufficiently strong attractions, bosons and fermions pair into molecules. Parallel to theoretical studies, boson-fermion Feshbach molecules were achieved also experimentally. The first realizations were obtained in Hamburg \cite{Ospelkaus_EXPBF3D_2006} and in Boulder \cite{Zirbel_ExpBF3D_2008} with ${}^{40}\textrm{K}\,-{}^{87}\textrm{Rb}$ mixtures. Later, the creation of Feshbach molecules was achieved also with an isotopic ${}^{40}\textrm{K}\,-{}^{41}\textrm{K}$ mixture \cite{Wu_ExpKK_2011}, as well as with ${}^{23}\textrm{Na}\,-{}^{6}\textrm{Li}$ \cite{Heo_Exp_BF3D_2012}, ${}^{23}\textrm{Na}\,-{}^{40}\textrm{K}$  \cite{Wu_Exp_3D_2012}, ${}^{87}\textrm{Rb}\,-{}^{40}\textrm{K}$  \cite{Cumby-2013}, , and ${}^{41}\textrm{K}\,-{}^{6}\textrm{Li}$ \cite{Fritsche-2021} heteronuclear BF mixtures. Recently, ${}^{23}\textrm{Na}\,-{}^{40}\textrm{K}$ Feshbach molecules have been successfully transferred to the absolute molecular ground state and cooled down to quantum degeneracy \cite{Schindewolf-2022}, while Ref.~\cite{Duda-2023} has investigated the same ${}^{23}\textrm{Na}\,-{}^{40}\textrm{K}$ mixture across the whole broad Feshbach resonance between bosons and fermions, confirming theoretical predictions \cite{Guidini_CondensedphaseBoseFermi_2015} about a universal behavior of the condensate fraction and the occurrence of a quantum phase transition.  
    
Equally interesting is the case of repulsive Bose-Fermi mixtures, which are characterized by both intra- and inter-species repulsive interactions. The formation of Feshbach molecules is here prohibited by the repulsive nature of the interactions. However, the problem of instability persists: sufficiently large repulsive boson-fermion interactions lead the system towards phase separation \cite{Roth_Stability_2002,Capuzzi_BF_mixture_2003,Titvinidze_Supersolid_2008,Lous_RepBFmixture_2018}.
    
The possibility of confining mixtures in lower dimensions using optical potentials elicited interest in two-dimensional (2D) mixtures. Besides the intrinsic interest of many-body systems in reduced dimensionality, the presence of a confining potential offers a further knob to tune the effective 2D interactions between the components of the mixture through a  confinement-induced resonance \cite{Olshanii-1998,Petrov-2000,Petrov-2001,Haller_ConfInducedResonance_2010}.
Interestingly, it has been shown in Ref.~\cite{Bazak_StableWaveResonant_2018} that such a mechanism in 2D could lead to the creation of collisionally stable fermionic dimers made by one boson and one fermion,  with a strong p-wave mutual attraction that could support a p-wave superfluid of dimers.
Besides this recent important result, relatively few theoretical studies have been conducted on 2D BF mixtures \cite{Fehrmann-2004,Wang_BF2D_2005,Mathey_BF2D_2006,Subasi-2010,Noda-2011,Von_Milczewski_BF2D_2022}.
    
This motivates the present paper, in which we present a combined effort of a (second-order) perturbative study and numerical non-perturbative quantum Monte Carlo (QMC) simulations of repulsive 2D Bose-Fermi mixtures, investigating the regime of validity of perturbation theory and the onset of phase separation. 

Our main results are: i) The derivation of the fermionic effective mass at the perturbative level and with a QMC estimation employing a novel finite-size extrapolation method; ii) The detailed analysis of the equation of state, with varying BF repulsion and bosonic concentration, for which we observe good agreement between perturbative and QMC results up to the predicted onset of phase separation; iii) The study of this onset from a stability condition viewpoint, and by inspection of QMC pair correlation functions.

The article is organized as follows. In Sec.~\ref{sec:perturbative} we develop the perturbative theory up to second order, deriving an expansion for the chemical potentials and equation of state, the effective mass, and the stability condition. In Sec.~\ref{sec:qmc} we describe the Monte Carlo methods that we used, including a new finite-size correction scheme for BF mixtures, and report our results for the effective mass, the equation of state, and the bosonic pair distribution function for different bosonic concentrations. Finally, Sec.~\ref{sec:conclusions} reports our conclusions. The appendices provide further details on the perturbative expansion and the QMC method.

\section{Perturbative expansion}\label{sec:perturbative}
We consider a 2D BF mixture with fermionic particle density $n_{\rm F}$ and bosonic particle density $n_{\rm B} = x n_{\rm F}$, where $x$ is the bosonic concentration. The atomic masses are $m_{\rm F}$ and $m_{\rm B}=w m_{\rm F}$, respectively, where we introduced the mass ratio $w$. We are interested in developing a perturbative treatment for Bose-Fermi mixtures. Previous perturbative works have indeed studied Bose-Fermi mixtures only in three dimensions \cite{Albus-2002,Viverit-2002} while only Fermi-Fermi \cite{Bloom-1975,Bruch-1978,Engelbrecht-1992,Chubukov-2010,Anderson-2011,Beane_precisionFermiliquidtheory_2023} or Bose-Bose systems \cite{Schick-1971,Popov-1972,Cherny-2001,Andersen-2002,Mora-2003,Mora_GroundStateEnergy_2009} have been considered in two dimensions.
Our perturbative treatment will be valid for generic mass-imbalanced mixtures, for both attractive and repulsive BF interactions, while we will focus on the repulsive case with equal masses ($w=1$) in the QMC simulations.
\subsection{Diluteness condition and expansion parameters}
We assume the system to be {\em dilute}, such that the average distance between any pair of particles of the mixture is much larger than the range $R$ of their interaction. Under this condition, the boson-fermion and boson-boson interactions can be parametrized in terms of the corresponding (2D) $s$-wave scattering lengths $a_{\rm BF}$ and $a_{\rm BB}$, while fermion-fermion interactions  can be altogether neglected since $s$-wave interactions between identical fermions are forbidden by Fermi statistics and direct $p$-wave (or higher angular momenta) interactions are strongly suppressed.  

Specifically, for a generic finite-range two-body interaction, the $s$-wave phase shift, which yields the dominant contribution to the scattering amplitude at low relative momenta $k$, has the following effective-range expansion in 2D for $k$ approaching zero (see, e.g., \cite{Averbuch-1986,Hammer_Causalityeffectiverange_2010}):
\begin{equation}\label{eq:phaseshift}
    \cot{\delta_0(k)} = \frac{2}{\pi}\ln({k a})+O(k^2)
\end{equation}
where the length $a$ appearing within the logarithm defines the $s$-wave scattering length, and the constant finite terms in the limit $k\to 0$ have been included in its definition. With this convention, the 2D scattering length of a hard-disk potential of radius $R$ is $a=e^\gamma R/2$, where $\gamma\simeq 0.577216$ is Euler-Mascheroni constant, while, in the attractive case, the dimer binding energy is $-\hbar^2/(m_r a^2)$, where $m_r$ is the reduced mass. An alternative convention (used for example in Refs.~\cite{Astrakharchik_Lowdimensionalweaklyinteracting_2010,Bertaina_BCSBECCrossoverTwoDimensional_2011,Bertaina_Twodimensionalshortrangeinteracting_2013,Galea2017}) would correspond to $a=R$ for a hard-disk potential.

In the repulsive case, the scattering length $a$ is of the same order of the range $R$ of interaction (for a strong barrier) or even much smaller than $R$ (for a weak barrier). The diluteness condition then automatically implies that the gas parameters $n_{{\rm B}} a_{\rm BB}^2$ and $n_{\rm B,F}a_{\rm BF}^2$ are much smaller than 1.
Actually, in two dimensions, due to the logarithmic dependence of the scattering amplitude on the relative momentum and energy, it is convenient to describe boson-boson and boson-fermion interactions in terms of the dimensionless coupling parameters $g_{\rm BF}\equiv -1/\ln(k_{\rm F}a_{\rm BF})$ and $g_{\rm BB}\equiv -1/\ln(n_{\rm B}a_{\rm BB}^2)$, where the Fermi momentum $k_{\rm F}$ is related to the fermion density $n_{\rm F}$ by the equation $n_{\rm F}=k_{\rm F}^2/(4\pi)$.

For attractive BF interaction, the scattering length $a_{\rm BF}$ coincides with the bound-state radius, and is not necessarily related to the range $R$ of the interaction (the range $R$ can even be vanishing in this case). Perturbation theory requires in this case $k_{\rm F}a_{\rm BF} \gg 1$, corresponding to a weakly-bound two-body bound state with a large radius compared with the average interparticle distance and implying that  $g_{\rm BF} = -1/\ln(k_{\rm F}a_{\rm BF})$ is small and negative. The BF system is thus dilute with respect to the range  ($k_{\rm F} R 
\ll 1$) but dense with respect to the bound state radius ($k_{\rm F}a_{\rm BF} \gg 1$). These differences notwithstanding, perturbation theory is formally identical in the two cases.

In both cases, our perturbative expansion is constructed by considering $g_{\rm BF}$ and $g_{\rm BB}$ of the same order, say  $g_{\rm BF}= \alpha_{\rm BF} g$ and $g_{\rm BB}= \alpha_{\rm BB} g$, where $\alpha_{\rm BF}$ and $\alpha_{\rm BB}$ are some numerical constant, and taking the limit $g\to 0$. We will be interested in particular in deriving a perturbative expansion to second order in the small parameter $g$.

For brevity, in the rest of this section we set $\hbar=1$.

\subsection{Many-body T-matrix}
\begin{figure}[htbp]
\centering
\includegraphics[width=\columnwidth]{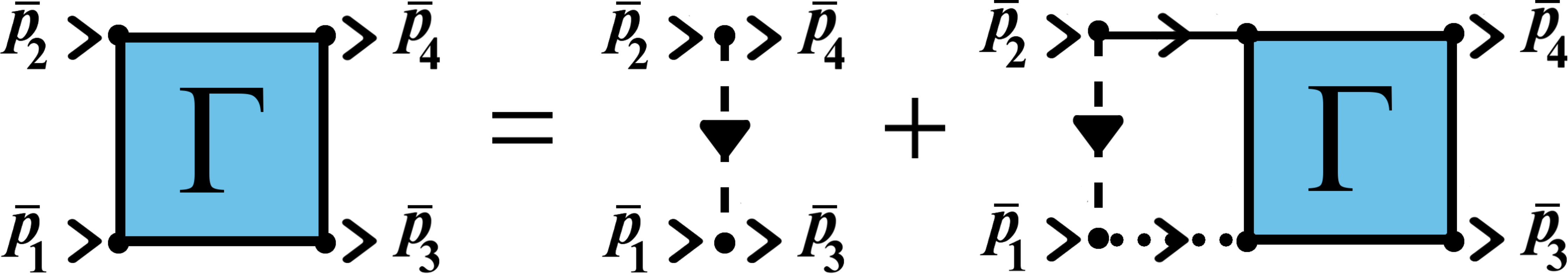}
\caption{Feynman diagrams for the boson-fermion many-body $T$-matrix  $\Gamma(\bar{p}_1,\bar{p}_2;\bar{p}_3,\bar{p}_4)$. Full lines correspond to bare fermion Green's functions $G_{\rm F}^0$, dotted lines to bare bosonic Green's functions $G_{\rm B}^0$, dashed lines to boson-fermions interactions $U$, while the blue box indicates the  boson-fermion many-body T-matrix  $\Gamma$. Arrows indicate the flow of momentum, which is conserved at each vertex (indicated by a dot). }  
\label{fig:tmatrix}
\end{figure}

The basic  building block of perturbation theory for a dilute Bose-Fermi mixture is the generalization to the many-body system of the two-body T-matrix. The many-body T-matrix $\Gamma(\bar{p}_1,\bar{p}_2;\bar{p}_3,\bar{p}_4)$ can be interpreted as a generalized scattering amplitude in the medium, accounting for the influence of the other particles  in the scattering processes. In terms of Feynman's diagrams, it is the sum of ladder diagrams (see Fig.\ref{fig:tmatrix})
and, in 2D, it corresponds to the following integral equation: 
\begin{eqnarray} \label{eq:bethesalpeter}
&&\Gamma(\bar{p}_1,\bar{p}_2;\bar{p}_3,\bar{p}_4)=V(\textbf{p}_\textbf{3}-\textbf{p}_\textbf{1})
+ i \int \!\!\frac{d\bar{p}}{(2\pi)^{3}}  V(\textbf{p}-\textbf{p}_\textbf{1})\nonumber \\ &&\times \; G_{\rm B}^0(\bar{p})G_{\rm F}^0(\bar{p}_1+\bar{p}_2-\bar{p})\Gamma(\bar{p},\bar{p}_1+\bar{p}_2-\bar{p};\bar{p}_3,\bar{p}_4).
\end{eqnarray}

Here, overbarred quantities indicate $(2+1)$-vectors: $\bar{p}=({\bf p},p_0)$, where ${\bf p}$ is a momentum variable and $p_0$  is a frequency;
 $V({\bf q})=\int d^2r e^{-i {\bf q}\cdot {\bf r}}V({\bf r})$ is the Fourier transform of the interaction potential $V({\bf r})$ between bosons and fermions at distance ${\bf r}$, while the bare boson and fermion Green's functions at zero temperature are given by 
\begin{eqnarray}
G_{\rm B}^0(\bar p)&=&\frac{1}{p_0 - p^2/2 m_{\rm B}  + \mu_{\rm B} + i \eta} \\  
G_{\rm F}^0(\bar p)&=&\frac{1}{p_0 - p^2/2 m_{\rm F} + i \eta \, {\rm sgn}(k-k_{\rm F})} ,
\end{eqnarray}
where $\eta$ is an infinitesimal positive quantity and $\mu_{\rm B}$ is the boson chemical potential.
Translational invariance and the instantaneous nature of the interaction potential imply that the many-body T-matrix $\Gamma$ depends on three momenta and one frequency.   In particular, by defining $\bar{P}=\bar{p}_1+\bar{p}_2=\bar{p}_3+\bar{p}_4, \mathbf{k}=(\mathbf{p}_1-\mathbf{p}_2)/2, \mathbf{k}'=(\mathbf{p}_3-\mathbf{p}_4)/2$, and integrating over the frequency $p_0$ in Eq.~(\ref{eq:bethesalpeter}), one obtains
\begin{eqnarray}
&&\Gamma(\textbf{k}',\textbf{k};\bar{P})=V(\textbf{k}'-\textbf{k})
	+\int \!\!\frac{d\textbf{p}}{(2\pi)^2} V\left(\textbf{k}-\textbf{p}\right)\nonumber\\
&&\times  \frac{ \Theta(|\textbf{P}/2-\textbf{p}|-k_{\rm F})\Gamma\left(\textbf{k}',\textbf{p};\bar{P}\right)}{ P_0 - \frac{\left(\textbf{P}/2-\textbf{p}\right)^2}{2m_{\rm F}} - \frac{\left(\textbf{P}/2+\textbf{p}\right)^2}{2m_{\rm B}}+\mu_{\rm B} + i\eta}.
\label{gamma}
\end{eqnarray}
For vanishing densities, such that $k_{\rm F}\to 0$, the above equation reduces to the integral equation for the two-body T-matrix $T^{\rm 2B}(\textbf{k}',\textbf{k};z)$ of the quantum theory of scattering [cf.~Eq.\ref{eq:c} of appendix \ref{app:scattering}]
calculated at $z=P_0-P^2/2M+\mu_{\rm B}+i\eta$, where $M=m_{\rm B} + m_{\rm F}$, and the reduced mass $m_r=m_{\rm B} m_{\rm F}/(m_{\rm B} + m_{\rm F})$. 

The on-shell two-body $t$-matrix, $t(\textbf{k}',\textbf{k})$, is instead obtained by calculating $T_{\rm 2B}(\textbf{k}',\textbf{k},z)$ for  $z={\bf k}^2/2m+i\eta$, such that
\begin{equation}
t(\textbf{k}',\textbf{k})=V(\textbf{k}'-\textbf{k})+\int\!\! \frac{d\textbf{p}}{(2\pi)^2}\frac{2m_r \, V(\textbf{k}'-\textbf{p}) t(\textbf{p},\textbf{k})}{{\textbf{k}}^2-\textbf{p}^2+i\eta}.
\label{tmat}
\end{equation}
In analogy with the 3D case, the similarity in the structure of  Eqs.~(\ref{gamma}) and (\ref{tmat}) allows one to replace $V$ with $t$ in the integral equation (\ref{gamma}) (see, e.g. \cite{AGD, Albus-2002}), yielding:
\begin{eqnarray} \label{gamma-fin}
&&\Gamma(\textbf{k}',\textbf{k};\bar{P})=t(\textbf{k}',\textbf{k})+\int\!\!\frac{ d\textbf{p}}{(2\pi)^2} t(\textbf{k}',\textbf{p}) \Gamma\left(\textbf{p},\textbf{k};\bar{P}\right) \nonumber\\ 
&&\times\left[  \frac{ \Theta(|\textbf{P}/2-\textbf{p}|-k_{\rm F})}{ P_0 - \frac{\left(\textbf{P}/2-\textbf{p}\right)^2}{2m_{\rm F}} - \frac{\left(\textbf{P}/2+\textbf{p}\right)^2}{2m_{\rm B}}+\mu_{\rm B} + i\eta}
-\frac{2 m_r}{{\textbf{k}}^2-\textbf{p}^2+i\eta} \right] ,
\end{eqnarray}
which is the starting point for our perturbative calculations.  In particular, a perturbative expansion in the small parameter $g_{\rm BF}$  (see appendix \ref{app-T-matrix} for details) shows that to second order in  $g_{\rm BF}$  the many-body T-matrix  $\Gamma$ depends only on the total 3-momentum $\bar{P}$, then yielding $\Gamma(\textbf{k}',\textbf{k};\bar{P})= \Gamma(\bar{P})$ with
\begin{equation}
 \Gamma(\bar{P})=\frac{\pi g_{\rm BF}}{m_r}\left[1+\frac{g_{\rm BF}}{2}F_{\Gamma}(\bar{P})\right] + o(g_{\rm BF}^2) ,
 \label{gamma-expanded}
\end{equation}
where the dimensionless function  $F_{\Gamma}(\bar{P})$ is defined by Eqs.~(\ref{FG1}) and (\ref{FG2}) of appendix \ref{app-T-matrix}. Note that the expansion \eqref{gamma-expanded} is valid both in the repulsive ($g_{\rm BF} > 0$) and attractive ($g_{\rm BF} < 0$) cases.

Equation~(\ref{gamma-expanded}) clearly shows that $\Gamma$ is the basic building block of the perturbative expansion  with respect to the boson-fermion coupling parameter $g_{\rm BF}$. 

\subsection{Fermionic self-energy and chemical potential}
\begin{figure}[tbp]
\centering
\vspace{0.1cm}
\includegraphics[scale=0.37]{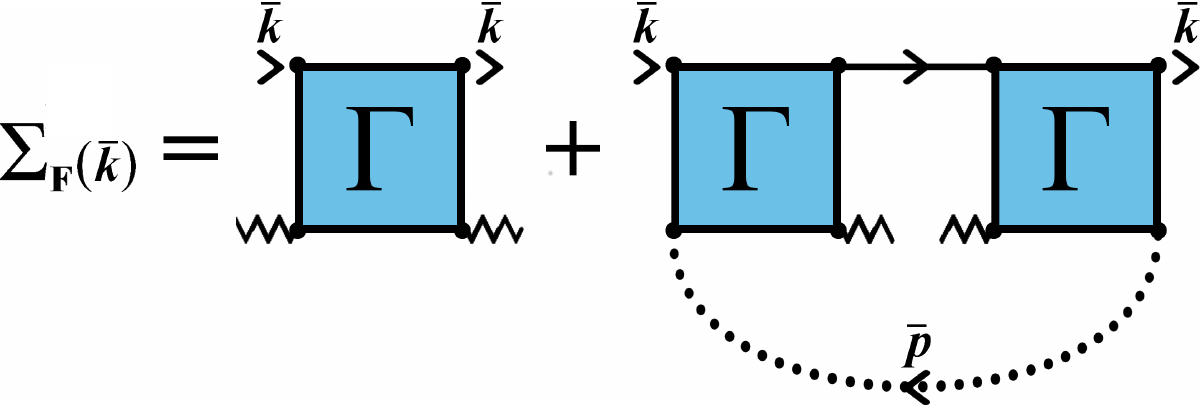}
\caption{Feynman diagrams for the fermionic irreducible self-energy to order $g^2$. Full lines correspond to bare fermion Green's functions $G_{\rm F}^0$, dotted lines to bare bosonic Green's functions, zig-zag lines correspond to factors $\sqrt{n_0}$, where $n_0$ is the condensate density, to be identified with the boson density $n_{\rm B}$ for consistency to order $g^2$.} 
\label{fig:SelfF}
\vspace{0.1cm}
\end{figure}

To second order in the small parameter $g$, only the two diagrams shown in Fig.~\ref{fig:SelfF} contribute to the irreducible fermionic self-energy $\Sigma_{\rm F}(k,\omega)$. It is indeed clear from Eq.~(\ref{gamma-expanded}) that diagrams containing three or more $\Gamma$ would contribute only to order $g^3$ or higher. In the same way, 
diagrams obtained by replacing in the first diagram of Fig.~\ref{fig:SelfF} the two condensate factors with a bosonic line dressed by self-energy insertions (including anomalous terms) contribute only higher order terms in the small parameter $g$, apart for a term of order $g_{\rm BF} g_{\rm BB}$ originating from the condensate depletion induced by the boson-boson interaction. This term is automatically taken into account by replacing $n_0$ with $n_{\rm B}$ in the diagrams of Fig.~\ref{fig:SelfF}.
Note further that the diagram obtained by the second diagram of Fig.~\ref{fig:SelfF} by replacing the dotted line with two condensate insertions is reducible and thus does not contribute to the irreducible self-energy.

In order to  calculate perturbatively the fermionic chemical potential $\mu_{\rm F}$, as well as the quasi-particle residue
 $Z(k_{\rm F})$ and fermionic effective mass $m^*$, we need $\Sigma_{\rm F}(k,\omega)$ in a neighborhood of $(k_{\rm F},\varepsilon_{\rm F})$, where $\varepsilon_{\rm F}=k_{\rm F}^2/2 m_{\rm F}=2\pi n_{\rm F}/m_{\rm F}$ is the Fermi energy in the absence of interactions. In this region, the self-energy $\Sigma_{\rm F}(k,\omega)$ is given by (see appendix \ref{app-selfF} for details):
\begin{widetext}
\begin{equation} \label{eq:SelfF2DCompleta}
\begin{split}
\Sigma_{\rm F}(k,\omega)=&\frac{\pi n_{\rm B} g_{\rm BF}}{m_r} + \frac{\pi n_{\rm B} g_{\rm BF}^2}{2m_r} \Biggl\{ 
 \ln \Biggl(\frac{(w+1)^2+A(w+1)-2\kappa^2}{(w+1)^2}- i\eta
+\sqrt{1+ \frac{2A}{w+1}-\frac{4\kappa^2-A^2}{(w+1)^2} -i\eta} \Biggr) -\ln 2\\
&+ \frac{w+1}{w-1} \ln \frac{B(w-1)-(w-1)^2+2\kappa^2-(w-1)\sqrt{(w-1-B)^2-4\kappa^2-i\eta}}{2\kappa^2} \Biggr\},
\end{split}
\end{equation}
\end{widetext}
where $A\equiv \kappa^2-\nu w$, $B\equiv \kappa^2+\nu w$, $\nu\equiv\omega/\varepsilon_{\rm F}$, and $\kappa \equiv k/k_{\rm F}$.

The chemical potential $\mu_{\rm F}$ is most easily obtained from the self-energy $\Sigma_{\rm F}(k,\omega)$ by using 
 Luttinger's theorem \cite{Luttinger-1960}, which connects the fermionic chemical potential to the real part of the self-energy calculated at the Fermi momentum $k_{\rm F}$ and  frequency $\omega=\mu_{\rm F}$:
\begin{equation} 
\label{Luttinger1}
\mu_{\rm F}= \varepsilon_{\rm F}+{\rm Re} \Sigma_{\rm F}(k=k_{\rm F},\omega=\mu_{\rm F}).
\end{equation}
 By noticing that the self-energy (\ref{eq:SelfF2DCompleta}) depends on the frequency only through the term proportional to $g_{\rm BF}^2$ and that $\mu_{\rm F} = \varepsilon_{\rm F} + O(g_{\rm BF})$, one sees that, neglecting terms of order higher than $g_{\rm BF}^2$,
 Eq.~(\ref{Luttinger1}) can be replaced with 
 \begin{equation} 
\label{Luttinger2}
 \mu_{\rm F}= \varepsilon_{\rm F}+{\rm Re} \Sigma_{\rm F}(k=k_{\rm F},\omega=\varepsilon_{\rm F}).
 \end{equation} 
By setting in (\ref{eq:SelfF2DCompleta}) $\kappa=1$ and $\nu=1$ (such that $A=1-w$ and $B=1+w$), one obtains
\begin{equation}
\Sigma_{\rm F}(k_{\rm F}, \varepsilon_{\rm F})=\frac{\pi n_{\rm B} g_{\rm BF}}{m_r} + \frac{\pi n_{\rm B} g_{\rm BF}^2}{2m_r} \Biggr[
 \ln \frac{w}{(w+1)^2} + \frac{w+1}{w-1} \ln w \Biggr],
\end{equation}
yielding
\begin{equation}
 \label{eq:muF-dimensionless}
\frac{\mu_{\rm F}}{\varepsilon_{\rm F}}=1 +\frac{w+1}{2w} x \, g_{\rm BF}^{}
\left( 1+ g_{\rm BF}^{} \ln \frac{w^{\frac{w}{w-1}}}{w+1} \right),
\end{equation}
where $x=n_{\rm B}/n_{\rm F}$ and we used $m_r = m_{\rm F} w/(w+1)$.
For equal masses, by taking the limit $w \to 1$, Eq.~(\ref{eq:muF-dimensionless}) reduces to
\begin{equation} \label{eq:muF-eq-masses}
\frac{\mu_{\rm F}}{\varepsilon_{\rm F}}=1+ x \, g_{\rm BF} \left(1+ g_{\rm BF}\ln \frac{e}{2}\right).
\end{equation}

\subsection{Ground-state energy and bosonic chemical potential}\label{subsec:energy}
The ground-state energy per unit volume ${\cal E}=E/V$ can be obtained by integrating the chemical potential $\mu_{\rm F}$ over the fermion density from 0 to $n_{\rm F}$:
\begin{equation}
{\cal E} = \int_0^{n_{\rm F}}\!\!\!\!\mu_{\rm F}(n'_{\rm F}) \, d n'_{\rm F} + {\cal E}_{\rm B}
\end{equation}
where ${\cal E}_{\rm B}$ is the ground-state energy per unit volume of the boson component in the absence of fermions.
We make the dependence of $\mu_{\rm F}$ on $n_{\rm F}$ in Eq.~(\ref{eq:muF-dimensionless}) more explicit 
\begin{equation}
 \label{eq:muF-n_F}
\mu_{\rm F}= \varepsilon_{\rm F}(n_{\rm F}) + c_1 g_{\rm BF}^{}(n_{\rm F}) + c_2 g_{\rm BF}^{2}(n_{\rm F})
\end{equation}
where $g_{\rm BF}=-1/\ln(k_{\rm F}a_{\rm BF})$ depends on $n_{\rm F}$ via $k_{\rm F}=\sqrt{4\pi n_{\rm F}}$
and
 $$c_1=\frac{\pi n_{\rm B}(w+1)}{m_{\rm F}\, w}, \phantom{aaaaaa} c_2=\frac{c_1}{2} \ln \frac{w^\frac{w}{ w-1}}{w+1}.$$
 Integration by parts of the terms depending on $g_{\rm BF}$ in Eq.~(\ref{eq:muF-n_F}) yields
\begin{eqnarray}
\int\!\!g_{\rm BF}^{} \, dn_{\rm F} &=&n_{\rm F}g_{\rm BF} -\frac{1}{2}\int\!\! g_{\rm BF}^2 \,dn_{\rm F}\\
\int\!\! g_{\rm BF}^2 \,dn_{\rm F} &=& n_{\rm F} g_{\rm BF}^2 -\int\!\! g_{\rm BF}^3 \,dn_{\rm F}=n_{\rm F} g_{\rm BF}^2 + o(g_{\rm BF}^2) .
\end{eqnarray}
Neglecting terms of order higher than $g_{\rm BF}^2$, we thus obtain
\begin{equation}\label{eq:EnergyPerturbative}
    {\cal E} = \frac{\varepsilon_{\rm F} n_{\rm F}}{2} + \frac{\varepsilon_{\rm F}n_{\rm B}(w+1)}{2 w}
    \left[g_{\rm BF} + g_{\rm BF}^2 \ln \frac{w^\frac{w}{ w-1}}{\sqrt{e}(w+1)}\right] 
 + {\cal E}_{\rm B}\;.
\end{equation}
We emphasize that, according to our expansion, Eq.~\eqref{eq:EnergyPerturbative} is valid to order $g^2$, that is, neglecting terms of order higher than two in $g_{\rm BF}$, $g_{\rm BB}$, or their combinations. Under this assumption, terms involving simultaneously $g_{\rm BF}$ and $g_{\rm BB}$ do not contribute to ${\cal E}$. 

In the mass-balanced case $m_{\rm B}=m_{\rm F}\equiv m$ on which we will focus in our QMC simulations, Eq.~~\eqref{eq:EnergyPerturbative} reduces to
\begin{equation}
\label{balanced}
 {\cal E} = \frac{\varepsilon_{\rm F} n_{\rm F}}{2} + \varepsilon_{\rm F}n_{\rm B}
    \left(g_{\rm BF}  - g_{\rm BF}^2 \ln \frac{2}{ \sqrt{e}} \right) 
 + {\cal E}_{\rm B}.
\end{equation}
We will verify, in particular, the validity of the term proportional to $ g_{\rm BF}^2$ in Eq.~\eqref{balanced}, which is the novel contribution to the equation of state derived by our perturbative calculation.

Concerning instead the bosonic term ${\cal E}_{\rm B}$, an accurate perturbative expansion has been obtained in \cite{Mora_GroundStateEnergy_2009}, and verified with QMC calculations shortly after \cite{Astrakharchik_Lowdimensionalweaklyinteracting_2010}. Its expression is 
\begin{equation}\label{eq:eosBosonsAstrakharchik}
   {\cal E}_{\rm B} = \frac{2\pi n_{\rm B}^2/m_{\rm B}}{\mathcal{L}+\ln{\mathcal{L}}+C_1^E+\frac{\ln{\mathcal{L}}+C_2^E}{\mathcal{L}}} \, ,
\end{equation}
where $C_1^E=-\ln\pi-2\gamma-1/2$, $C_2^E\simeq -0.05$ and $\mathcal{L}=g_{\rm BB}^{-1}-\ln{4}+2\gamma$ with our convention for the scattering length.
Neglecting terms of order higher than $g_{\rm BB}^2$, consistently with the order $g^2$ of our perturbative calculation, the above equation reads
\begin{equation}
    {\cal E}_{\rm B}  =\frac{2\pi n_{\rm B}^2}{m_{\rm B}} g_{\rm BB}\left[1+g_{\rm BB}\ln g_{\rm BB} + g_{\rm BB}\ln(4 \pi \sqrt{e})\right].
\end{equation}

Coming back to expression \eqref{eq:EnergyPerturbative} for the ground-state energy of the Bose-Fermi mixture, differentiation with respect to $n_{\rm B}$ yields immediately the bosonic chemical potential 
\begin{equation}
\label{mub}
 \mu_{\rm B} = \varepsilon_{\rm F} \frac{w+1}{2 w}
    \left[g_{\rm BF} + g_{\rm BF}^2 \ln \frac{w^\frac{w}{ w-1}}{\sqrt{e}(w+1)}\right] + \mu_{\rm B}^0,
\end{equation}  
where $\mu_{\rm B}^0 =\partial{{\cal E}_{\rm B}}/\partial n_{\rm B}$ is the boson chemical potential in the absence of fermions. Neglecting terms of order higher than $g_{\rm BB}^2$, it is given by
\begin{equation}
    \mu_{\rm B}^0  =\frac{4\pi n_{\rm B}}{m_{\rm B}} g_{\rm BB}\left[1+g_{\rm BB}\ln g_{\rm BB} + g_{\rm BB}\ln(4 \pi)\right].
\end{equation}

For equal masses, Eq. \eqref{mub} reduces to

\begin{equation}
 \mu_{\rm B} = \varepsilon_{\rm F}
    \left(g_{\rm BF}  - g_{\rm BF}^2 \ln \frac{2}{ \sqrt{e}} \right) + \mu_{\rm B}^0.
\end{equation}

\subsection{Quasi-particle residue}
The quasi-particle residue is obtained by the relation:
\begin{equation} 
Z(k_{\rm F})=\left[ 1- \frac{\partial}{\partial \omega} {\rm Re}\Sigma_{\rm F}(k_{\rm F},\omega)\right]_{\omega=\mu_{\rm F}}^{-1}.
\end{equation}
Since $\mu_{\rm F} =  \varepsilon_{\rm F}  + O(g_{\rm BF})$ and the self-energy (\ref{eq:SelfF2DCompleta}) depends on the frequency only
 through the term proportional to $g_{\rm BF}^2$, one would be tempted to evaluate the derivative with respect to $\omega$ in the above equation for $\omega = \varepsilon_{\rm F}$. Such a derivative is however {\em divergent} as $\omega \to \varepsilon_{\rm F}$, and so it is important to evaluate it at $\mu_{\rm F}$ rather than $\varepsilon_{\rm F}$.

Indeed, by setting $\kappa = 1$ (corresponding to $k= k_{\rm F}$) in Eq. \eqref{eq:SelfF2DCompleta}, and analyzing separately the behavior of 
${\rm Re}\Sigma_{\rm F}(k_{\rm F},\omega)$ for $\omega \to \varepsilon_{\rm F}^+$ and $\omega \to \varepsilon_{\rm F}^-$ one obtains the following asymptotic behavior for $\frac{\partial {\rm Re}\Sigma_{\rm F}(k_{\rm F},\omega)}{\partial \omega}$ when $\omega \to \varepsilon_{\rm F}$ (corresponding to $\nu \to 1$):
\begin{equation} \label{eq:SigmaPartialOmega2D}
\frac{\partial {\rm Re}\Sigma_{\rm F}(k_{\rm F},\omega)}{\partial \omega} \Biggr\vert_{\omega \simeq \varepsilon_{\rm F}} \!\!\simeq \frac{\pi n_{\rm B} g_{\rm BF}^2}{2m_r \varepsilon_{\rm F} } \left( \frac{w+1}{w}-\frac{w+1}{2\sqrt{w|\nu-1|}}\right),
\end{equation}
which clearly shows that $\partial {\rm Re}\Sigma_{\rm F}(k_{\rm F},\omega)/\partial \omega$ diverges as $1/\sqrt{|\omega - \varepsilon_{\rm F}|}$ for $\omega\to\varepsilon_{\rm F}$.

By inserting expression \eqref{eq:muF-dimensionless} for $\mu_{\rm F}$ in Eq.~\eqref{eq:SigmaPartialOmega2D} and neglecting terms $o(g_{\rm BF}^2)$, we obtain
\begin{align}
Z(k_{\rm F}) = & \left[ 1- \frac{\pi n_{\rm B} g_{\rm BF}^2}{2m_r\varepsilon_{\rm F}} \left( \frac{w+1}{w}-\frac{w+1}{2\sqrt{w|\nu-1|}}\right) \right]_{\omega=\mu_{\rm F}}^{-1} \notag \\[4mm]
 \simeq & \left[ 1- \frac{\pi n_{\rm B} g_{\rm BF}^2}{2m_r \varepsilon_{\rm F}} \frac{w+1}{w} \left( 1 - \frac{1}{2} \sqrt{\frac{w m_r \varepsilon_{\rm F}}{ \pi n_{\rm B} |g_{\rm BF}|}} \right) \right]^{-1} \notag \\[4mm]
\simeq & 1-\frac{\sqrt{2}}{8}\frac{(w+1)^{3/2}}{w}|g_{\rm BF}|^{3/2}\sqrt{x}+\frac{1}{4}\left( \frac{w+1}{w} \right)^2 g_{\rm BF}^2 x, \label{eq:RenormalizationConstant2D}
\end{align}
where in the last line we have used $m_r= m_{\rm F}w/(w+1) $ and $m_{\rm F}\varepsilon_{\rm F}=2\pi n_{\rm F}$. Note that due to the divergence of $\partial {\rm Re} \Sigma_{\rm F}(k_{\rm F},\omega) / \partial \omega$ for $\omega\to\varepsilon_{\rm F}$, the leading term in the expansion for $Z(k_{\rm F})$ is proportional to $|g_{\rm BF}|^{3/2}$ rather than $g_{\rm BF}^2$, as one would get in the absence of such a divergence (and as it occurs e.g. in a Fermi-Fermi system \cite{Bloom-1975,Engelbrecht-1992}).  

Mathematically, this diverging derivative originates from the presence of a Fermi step function in the integral over momentum yielding the second-order term for $\Sigma_{\rm F}$.   In a Fermi-Fermi system this divergence is smeared by a further integral over  momentum and frequency, which is absent in the BF mixture because a fermionic line with arbitrary momentum and frequency is replaced with a condensate line with vanishing momentum and frequency.

\subsection{Effective mass}
The effective mass is obtained by the relation:
\begin{equation}
\label{defeffectivemass} 
\frac{m_{\rm F}}{m^*}=\left[ 1+\frac{m_{\rm F}}{k_{\rm F}}\frac{\partial {\rm Re}\Sigma_{\rm F}(k,\varepsilon_{\rm F})}{\partial k} \right]_{k=k_{\rm F}} Z(k_{\rm F}).
\end{equation}
By setting $\nu = 1$ in Eq. \eqref{eq:SelfF2DCompleta} and taking the derivative with respect $k$ of ${\rm Re}\Sigma_{\rm F}(k,\varepsilon_{\rm F})$, one obtains
\begin{equation} \label{eq:SigmaPartialK2D}
\frac{\partial {\rm Re} \Sigma_{\rm F}(k,\varepsilon_{\rm F})}{\partial k} \Biggr\vert_{k=k_{\rm F}}=-\frac{\pi n_{\rm B} g_{\rm BF}^2}{2m_r k_{\rm F}} \frac{2}{w},
\end{equation}
such that, by inserting Eqs.~\eqref{eq:SigmaPartialK2D} and \eqref{eq:RenormalizationConstant2D} in Eq.~\eqref{defeffectivemass}:
\begin{align}
\frac{m_{\rm F}}{m^*}= & \left[ 1 - \frac{w+1}{4w^2} g_{\rm BF}^2 x\right] \notag \\[2mm]
\times & \left[ 1-\frac{\sqrt{2}}{8}\frac{(w+1)^{3/2}}{w}|g_{\rm BF}|^{3/2}\sqrt{x}+\frac{1}{4} \left( \frac{w+1}{w} \right)^2  g_{\rm BF}^2 x\right] \notag \\[5mm]
= & \; 1 - \frac{\sqrt{2}}{8}\frac{(w+1)^{3/2}}{w}|g_{\rm BF}|^{3/2}\sqrt{x} + \frac{w+1}{4w} g_{\rm BF}^2 x , \label{eq:EffectiveMass3D}
\end{align}
which, for equal masses ($w=1$), reads:
\begin{equation}\label{eq:perturbativemass}
\frac{m}{m^*}= 1-\frac{\sqrt{x}}{2} |g_{\rm BF}|^{3/2} + \frac{x}{2} g_{\rm BF}^2 .
\end{equation}
Note again that the leading term in the expansion for $m^*$ is proportional to $|g_{\rm BF}|^{3/2}$ rather than $g_{\rm BF}^2$, a behavior which is directly inherited from the quasi-particle residue $Z(k_{\rm F})$.

\subsection{Mechanical stability}

Mechanical stability requires the compressibility matrix  $\partial n_i / \partial \mu_j$ (with $i = {\rm B, F}$) or, equivalently, its inverse $\partial\mu_i / \partial n_j$, to be positive definite \cite{Viverit_Stability_2000}. It corresponds to
\begin{equation}\label{eq:Stability1}
    \frac{\partial \mu_i}{\partial n_i} \geq 0 ,
\end{equation}
and
\begin{equation}\label{eq:Stability2}
    \frac{\partial \mu_{\rm F}}{\partial n_{\rm F}}\frac{\partial \mu_{\rm B}}{\partial n_{\rm B}} - \frac{\partial \mu_{\rm F}}{\partial n_{\rm B}}\frac{\partial \mu_{\rm B}}{\partial n_{\rm F}} \geq 0.
\end{equation}
The second inequality, using the symmetry of the compressibility matrix, reads
\begin{equation}\label{eq:Stability2bis}
    \frac{\partial \mu_{\rm F}}{\partial n_{\rm F}}\frac{\partial \mu_{\rm B}}{\partial n_{\rm B}} \geq  \left(\frac{\partial \mu_{\rm F}}{\partial n_{\rm B}}\right)^2,
\end{equation}
which, when satisfied, makes the inequality \eqref{eq:Stability1} to be respected automatically in both cases ($i = {\rm B, F}$) when it is verified for just one of them.
One has (always neglecting higher order terms)
\begin{align}\label{eq:dmuFdnF}
    \frac{\partial \mu_{\rm F}}{\partial n_{\rm F}} &= \frac{2\pi}{m_{\rm F}}\left(1+\frac{w+1}{4 w} x g_{\rm BF}^2\right)\\
\label{eq:dmuFdnB}
    \frac{\partial \mu_{\rm F}}{\partial n_{\rm B}} &= \frac{2\pi}{m_{\rm F}}\frac{w+1}{2w}  g_{\rm BF}\left( 1+ g_{\rm BF}^{} \ln \frac{w^\frac{w}{w-1}}{w+1}  \right)\\
  \label{eq:dmuBdnB}  
     \frac{\partial \mu_{\rm B}}{\partial n_{\rm B}} &=\frac{4\pi}{m_{\rm B}} g_{\rm BB}\left[1+g_{\rm BB}\ln g_{\rm BB} + g_{\rm BB}\ln(4 \pi e)\right].
\end{align}
One sees that $\partial \mu_{\rm F}/\partial n_{\rm F} > 0$ always. We thus need to satisfy only \eqref{eq:Stability2bis} for the stability of the Bose-Fermi mixture. When using  Eqs.~\eqref{eq:dmuFdnF}$-$\eqref{eq:dmuBdnB}, it reads
\begin{align}
\label{condition_full}
g^{}_{\rm BB}\left[1+g^{}_{\rm BB}\ln(4 \pi e g^{}_{\rm BB})\right] &\ge \frac{(w+1)^2 g_{\rm BF}^2 \left( 1+ g_{\rm BF}^{} \ln \frac{w^\frac{w}{w-1}}{w+1}  \right)^2}{8w \left(1+ \frac{w+1}{4 w} x g_{\rm BF}^2\right)}
\end{align}
which, for equal masses, reduces to
\begin{align}
\label{condition_full_equal}
g^{}_{\rm BB}\left[1+g^{}_{\rm BB}\ln(4 \pi e g^{}_{\rm BB})\right] &\ge \frac{g_{\rm BF}^2 \left( 1+ g_{\rm BF}^{} \ln \frac{e}{2}  \right)^2}{2+ x g_{\rm BF}^2}.
\end{align}
 The condition  \eqref{condition_full} can be simplified if one keeps only the leading order terms, corresponding to a ``mean-field" treatment of the BB and BF  interaction. At this level of approximation one obtains 
\begin{align}
\label{condition_MF}
g_{\rm BB}\ge \frac{(w+1)^2}{8w} g_{\rm BF}^2,
\end{align}
which, for equal masses, reduces to
\begin{align}
\label{condition_MF_eq_masses}
g_{\rm BB}\ge \frac{1}{2} g_{\rm BF}^2.
\end{align}
The condition for stability \eqref{condition_MF} is the counterpart for a 2D Bose-Fermi mixture of the analogous condition 
for stability in 3D obtained in \cite{Viverit_Stability_2000}, which can be written as 
\begin{align}
\label{condition_MF_3D}
k_{\rm F} a_{\rm BB}\ge \frac{(w+1)^2}{ 2 \pi w} (k_{\rm F} a_{\rm BF})^2.
\end{align}
Both condition \eqref{condition_MF} in 2D and condition \eqref{condition_MF_3D} in 3D can be interpreted as the requirement that the direct BB repulsion overcomes the effective attraction between bosons induced by interactions with fermions, in such a way that the overall effective interaction between bosons remains repulsive \cite{Xin-2024}:
\begin{equation}
\label{induced}
{\cal T}_{\rm BB} + {\cal T}_{\rm BF}^2\left(-\frac{\partial n^{(0)}_{\rm F}}{\partial \mu^{(0)}_{\rm F}}\right) > 0.
\end{equation}
Here
${\cal T}_{\rm BB}$ and ${\cal T}_{\rm BF}$ are the leading order expression in the weak-coupling limit of the T-matrices for BB and BF interactions, and are given by 
\begin{equation}
\label{t2d}
    {\cal T}_{\rm BB} = \frac{4 \pi g_{\rm BB}}{m_{\rm BB}}, \phantom{aaaaaa}  {\cal T}_{\rm BF}  = \frac{\pi g_{\rm BF}}{m_r}
\end{equation}
in 2D and by 
\begin{equation}
\label{t3d}
    {\cal T}_{\rm BB} = \frac{4 \pi a_{\rm BB}}{m_{\rm BB}}, \phantom{aaaaaa}  {\cal T}_{\rm BF}  = \frac{2 \pi a_{\rm BF}}{m_r}
\end{equation}
in 3D, while $\frac{\partial n^{(0)}_{\rm F}}{\partial \mu^{(0)}_{\rm F}}$ is the compressibility of the ideal Fermi gas, which is given by
\begin{equation}
\label{compr2d}
    \frac{\partial n^{(0)}_{\rm F}}{ \partial \mu^{(0)}_{\rm F}} = \frac{2\pi}{m_{\rm F}}
\end{equation}
in 2D and by 
\begin{equation}
\label{compr3d}
    \frac{\partial n^{(0)}_{\rm F}}{ \partial \mu^{(0)}_{\rm F}} = \frac{m_{\rm F} k_{\rm F}}{2\pi^2}
\end{equation}
in 3D. It is straightforward to verify that conditions \eqref{condition_MF} 
and  \eqref{condition_MF_3D} are obtained from condition \eqref{induced} when Eqs.~\eqref{t2d}, \eqref{compr2d} or \eqref{t3d}, \eqref{compr3d} are used, respectively.

\section{Quantum Monte Carlo}\label{sec:qmc}

\subsection{Method}
To determine the ground state properties of a repulsive 2D Bose-Fermi mixture, we employ two different QMC techniques: variational Monte Carlo (VMC) and diffusion Monte Carlo (DMC). The VMC method stems from the application of Monte Carlo integration to the evaluation of quantum expectation values in chosen and suitably optimized variational trial wavefunctions $\psi_T$, and it is designed to be directly applied to both bosonic and fermionic systems~\cite{kalos_helium_1974,CeperleyMonteCarlosimulation1977}. Diffusion Monte Carlo is instead a more sophisticated technique that allows  for a stochastic solution of the many-body Schr\"{o}dinger equation in imaginary time. It is in principle an exact method for bosonic systems, but in the presence of fermions the sign problem arises. We use the standard fixed-node approximation, which consists in imposing that the nodal surface of the true many-body wave function is the same as the one of the trial wave function $\psi_T$ employed in our VMC simulations \cite{Ceperley_GroundStateElectron_1980}. Thus, both techniques provide an upper bound to the true ground state energy \cite{Reynolds_FNDMC_1982}, which can be lowered by variationally optimizing $\psi_T$.

The considered system is described by an effective low-energy Hamiltonian, which is set to reproduce the scattering lengths of the full atomic problem. Its expression is the following:
\begin{equation}
    \begin{split}
        H = &-\frac{\hbar^2}{2m_{\rm F}}\sum_{i=1}^{N_{\rm F}}\nabla_i^2 - \frac{\hbar^2}{2m_{\rm B}}\sum_{i'=1}^{N_{\rm B}}\nabla_{i'}^2 \\
        &+\sum_{i,i'=1}^{N_{\rm F}, N_{\rm B}}V_{\rm BF}(r_{ii'}) + \sum_{i'<j'}^{N_{\rm B}}V_{\rm BB}(r_{i'j'}),
    \end{split}
\end{equation}
where $i,j,...$ and $i',j',...$ label, respectively, fermions and bosons, $N_{\rm F}$ and $N_{\rm B}$ are their numbers, and $r_{kl}$ is the distance between particles $k$ and $l$. The short-range fermion-fermion interaction can be neglected at low energies due to the Pauli exclusion principle, while the specific form of the interaction potentials, $V_{\rm BF}(r)$ and $V_{\rm BB}(r)$, is irrelevant in the dilute regime of interest for ultracold gases. In particular, we assume a soft-disk potential for both interactions: $V_{\rm BB}(r)=V_{\rm BB}^0$ for $r<R_{\rm BB}$ and zero elsewhere, and similarly for $V_{\rm BF}(r)$, and parametrize the strength of the interactions $V_{\rm BB}^0$ and $V_{\rm BF}^0$ in terms of their respective scattering lengths, according to the relation $a_{\rm P}/(e^{\gamma}R_{\rm P}/2)=\exp{\left[-I_0(\zeta_{\rm P}R_{\rm P})/(\zeta_{\rm P}R_{\rm P} I_1(\zeta_{\rm P}R_{\rm P}))\right]}$~\cite{Pilati_QuantumMonteCarlo_2005} where the subscript ${\rm P}$ indicates the ${\rm BB}$ or ${\rm BF}$ pair, $\zeta_{\rm P}^2=2m_{\rm P} V_{\rm P}^0/\hbar^2$, $m_{\rm P}$ is the reduced mass of the ${\rm P}$ pair and $I_n$ is the modified Bessel function of order $n$. In this work, we tune both $V_{\rm BB}^0$ and $V_{\rm BF}^0$ so that the right-hand side of the former relation is equal to $1/2$. The results for the observables depend only on the scattering lengths, provided that the bare radii of the model potentials, $R_{\rm BB}$ and $R_{\rm BF}$, are negligible when compared to the densities of the two components: $n_{\rm F} R_{\rm BF}^2\ll 1$ and $n_{\rm B} R_{\rm BB}^2\ll 1$. For the BB radius, we have set $n_{\rm B} R_{\rm BB}^2 =x n_{\rm F} R_{\rm BB}^2=10^{-6} x$. The value of $n_{\rm F} R_{\rm BF}^2$ changes according to $g_{\rm BF}$ and is smaller than $10^{-3}$ in the homogeneous phase.

Simulations are carried out in a square box of area $L^2=N_{\rm F}/n_{\rm F}$ with periodic boundary conditions (PBC), with a number of fermions up to $N_{\rm F}=81$ and a number of bosons $N_{\rm B} < N_{\rm F}$ depending on the targeted bosonic concentration $x$. The sizes of our simulation boxes are quite large, as compared to the relevant kinetic length scale $k_{\rm F}^{-1}$, since $k_{\rm F} L=(4\pi N_{\rm F})^{1/2}$ which is $\sim 25$ for $N_{\rm F}=49$ or $\sim 32$ for $N_{\rm F}=81$. We focus on the case of equal masses, thus $m_{\rm B}=m_{\rm F}=m$. We use a Jastrow-Slater trial wavefunction, which turned out to be a good ansatz in the three-dimensional case \cite{Lee_MonteCarlostudy_1981,Krotscheck_Theory3He4Hemixtures_1993,Guidini_CondensedphaseBoseFermi_2015}. Thus, $\psi_T(\bm{R})$ is given by the product of two terms: $\Phi_S(\bm{R})$ and $\Phi_A(\bm{R})$. $\Phi_S(\bm{R})$ is a function of the particle coordinates $\bm{R}=(\bm{r}_1,...,\bm{r}_{N_{\rm F}},\bm{r}_{1'},...,\bm{r}_{N_{\rm B}})$ which is symmetric upon exchange of any two fermions or two bosons, and contains information regarding the boson-boson and the boson-fermion correlations. In the Jastrow form, this term reads as $\Phi_S(\bm{R})=\prod_{i,i'}f_{\rm BF}(r_{i,i'})\prod_{i'<j'}f_{\rm BB}(r_{i'j'})$, where the functions $f$ describe the two-body correlations and are solutions of the two-body problem with suitable boundary conditions. In particular, PBC require null first derivative of $f$ at distance $r=L/2$. For this reason, we introduce two variational parameters $\bar{R}_{\rm BB}\le L/2$ and $\bar{R}_{\rm BF}\le L/2$, to be optimized, that allow one to parametrize the distance at which the two-body Jastrow correlations go to a constant, with null first derivative~\cite{Pilati_QuantumMonteCarlo_2005,Yu_Stability_2011}. See appendix ~\ref{app:wavefunction} for the explicit form of Jastrow correlations. The second term, $\Phi_A(\bm{R})$, satisfies the fermionic antisymmetry condition and determines the nodal surface of $\psi_T$. In particular, we use a Slater determinant made of single particle orbitals in the form of plane waves $\exp{(i \bm{k} \cdot \bm{r}_i})$, whose wavevectors are $\bm{k}=(n_x,n_y)2\pi/L$, where $n_x$ and $n_y$ are integer numbers. As customary, these wavevectors are chosen so as to fill closed shells, which are closed with respect to mirror and discrete rotational symmetries, in order to reduce finite-size effects~\cite{Lin_Twistaveragedboundaryconditions_2001}. The optimized $\psi_T$ is then obtained by choosing the pairs of parameters $\bar{R}_{\rm BB}$ and $\bar{R}_{\rm BF}$ so as to minimize the VMC ground state energy.

\subsection{Finite-size analysis}\label{subsec:finitesize}
In order to make simulations computationally affordable, it is necessary to limit the study to systems with a relatively small number of particles. This implies inaccuracies in the QMC results, due to finite-size effects, with respect to the thermodynamic limit. To reduce them, we extend a finite-size correction originally developed in the case of Fermi mixtures \cite{Tanatar_Groundstatetwodimensional_1989,Kwon_QuantumMonteCarlo_1994,Bertaina_QuantumMonteCarlo_2023} to the case of Bose-Fermi mixtures. We assume that the finite-size correction related to the purely bosonic component is negligible, since we focus on relatively small bosonic concentrations, and the fermionic finite-size effects are presumably much more relevant due to shell effects.

\begin{figure}[tbp]
    \centering
    \includegraphics[width=\columnwidth]{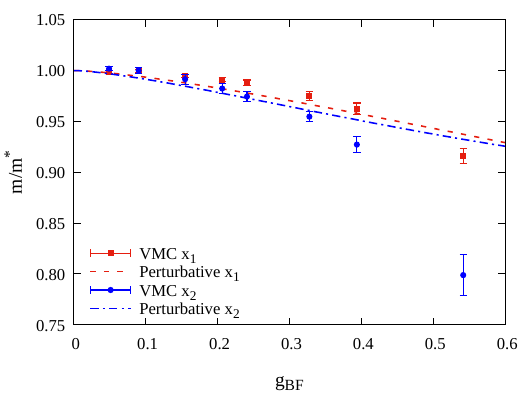}
    \caption{Inverse effective mass $m/m^*$ as a function of the BF coupling parameter for two different bosonic concentrations $x_1\simeq0.245$ and $x_2\simeq0.490$ and bosonic repulsion $g_{\rm BB} \simeq 5.9  \cdot 10^{-2}$ and $6.2  \cdot 10^{-2}$, respectively. The dashed (red) line and the dot-dashed (blue) curve represent the perturbative prediction Eq.~\eqref{eq:perturbativemass} using $x_1$ and $x_2$, respectively, while the symbols are the result of the finite-size correction procedure from VMC data.}
    \label{fig:EffectiveMass}
\end{figure}

We consider that the fermions do indirectly interact via the mediation of bosons. This allows us to use Landau Fermi liquid theory to elaborate an extrapolation scheme to the thermodynamic limit. 
In this approach, we assume that the main finite-size correction is analogous to the one of a noninteracting Fermi system, which is a purely kinetic energy contribution, introducing the equation:
\begin{equation}
    \varepsilon(n_{\rm F},N_{\rm F},N_{\rm B})=\varepsilon_{\infty}(n_{\rm F},x)-b\Delta t(n_{\rm F},N_{\rm F}),
\end{equation}
where $\varepsilon(n_{\rm F},N_{\rm F},N_{\rm B})$ and $\varepsilon_{\infty}(n_{\rm F},x)$ are the energies per fermion of the finite system with PBC and of the infinite system with the same fermion density $n_{\rm F} = N_{\rm F}/L^2$ and boson concentration $x=n_{\rm B}/n_{\rm F}=N_{\rm B}/N_{\rm F}$, respectively, while $\Delta t(n_{\rm F},N_{\rm F})=\sum_{\bm{k}}\hbar^2|\bm{k}|^2/(2m_{\rm F} N_{\rm F})-\varepsilon_{\rm F}/2$ is the energy difference per fermion between a system of $N_{\rm F}$ noninteracting fermions in the same square box with PBC (the sum is restricted to the wavevectors considered in the Slater determinant) and the infinite system with the same fermion density $n_{\rm F}$, which is easily tabulated~\cite{Lin_Twistaveragedboundaryconditions_2001}. The parameter $b$ can be identified as the inverse of the effective mass $m/m^*$ and can be determined by fitting the QMC data obtained for different values of $N_{\rm F}$, in analogy with the Fermi mixture case \cite{Tanatar_Groundstatetwodimensional_1989,Kwon_QuantumMonteCarlo_1994,Bertaina_QuantumMonteCarlo_2023}. Notice that, since interaction in this Fermi liquid is mediated by bosons, the term $\varepsilon_{\infty}(n_{\rm F},x)$ depends both on the fermionic density $n_{\rm F}$ and on the bosonic concentration $x$. Unfortunately, it is not possible to perform simulations with different $N_{\rm F}$ and $N_{\rm B}$, which vary discretely, while keeping constant $x$. Thus, simulations with different $N_{\rm F}$ also correspond to different $x$, which we are not free to consider as an independent variable. This prompts us to expand $\varepsilon_{\infty}(n_{\rm F},x)$ around a reference bosonic concentration $x_i$ in terms of the variation $x-x_i$. We therefore introduce the following scaling equation:
\begin{equation}\label{eq:FitEffectiveMass}
    \varepsilon(n_{\rm F},N_{\rm F},N_{\rm B})=\frac{\varepsilon_{\rm F}}{2}+\varepsilon_{\rm B}(n_{\rm B})x+\varepsilon_i+c_i(x-x_i)-b_i\Delta t({n_{\rm F},N_{\rm F}})\;,
\end{equation}
where the first term is the contribution of a gas of noninteracting fermions in the thermodynamic limit, while $\varepsilon_{\rm B}(n_{\rm B}) ={\cal E}_{\rm B}/ n_{\rm B}$ is the energy per boson of the corresponding interacting system of bosons in the absence of fermions, for which we can use the accurate equation of state \cite{Astrakharchik_Lowdimensionalweaklyinteracting_2010} given by expression \eqref{eq:eosBosonsAstrakharchik}. Since we assume that finite size effects of the purely bosonic contribution are negligible with respect to the fermionic ones, the energy per boson $\varepsilon_{\rm B}$ is also taken in the thermodynamic limit. The remaining parameters $\varepsilon_i$, $c_i$ and $b_i$ are obtained by fitting QMC results and model the residual correlation energy in the vicinity of $x\simeq x_i$. The above functional form is not the only possible choice for correcting for finite-size effects. In fact, we tested alternative expansions, for example by introducing contributions explicitly depending on inverse powers of $N_{\rm F}$, and we realized that the parameter $b_i$ is quite stable with respect to the different fitted functional forms, while Eq.~\eqref{eq:FitEffectiveMass} allows us to obtain reasonable values for the reduced chi-square. Nevertheless, in consideration of this variability in compatible fitting models, we have increased the uncertainty bars for $b_i$ at the level of $10^{-3}$, for interactions $g_{\rm BF}\gtrsim 0.25$.

We fitted the above scaling equation to VMC data, obtaining the effective mass and thus a finite-size correction for each considered BF interaction value, that we applied to correct VMC and DMC results using the expression $\varepsilon_{\infty}(n_{\rm F},x)=\varepsilon(n_{\rm F},N_{\rm F},N_{\rm B})+b_i\Delta t(n_{\rm F},N_{\rm F})$. See App.~\ref{app:fit} for an example of such a fitting procedure. For some test case we applied the same fitting technique also to DMC data, and observed that the resulting effective masses were compatible. We attribute this to a prominent role of the nodal surface in determining the effective mass. Previous studies on the electron gas indeed showed an important role of backflow correlations in the Slater determinant \cite{Kwon_QuantumMonteCarlo_1994,Azadi_QuasiparticleEffectiveMass_2021}. In the context of BF mixtures, this suggests a relevant future research direction, together with the comparison to other effective-mass calculation approaches within QMC methods, such as the evaluation of imaginary-time diffusion of quasiparticles~\cite{Boninsegni_PathIntegralMonte_1995}, the explicit simulation of finite-momentum wavefunctions~\cite{AriasdeSaavedra_Effectivemassone_1994}, or the  methodology recently introduced in \cite{Holzmann-2023} based on the calculation of the static self-energy.

\begin{figure}[tbp]
    \centering
    \includegraphics[width=\columnwidth]{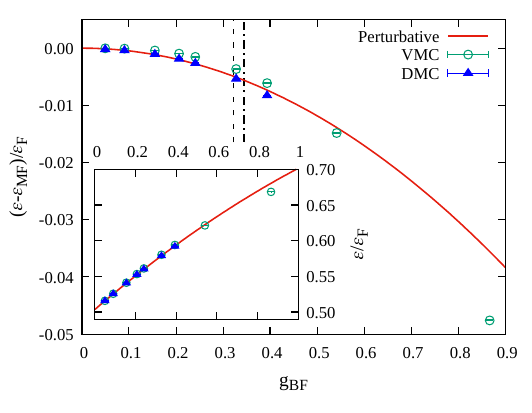}
  \caption{Zero-temperature equation of state for the bosonic concentration $x_1 \simeq 0.245$ and bosonic repulsion $g_{\rm BB} \simeq 5.9  \cdot 10^{-2}$, from VMC (empty circles) and DMC (filled squares). The energy per fermion, minus the mean-field term, is plotted as a function of the BF coupling parameter. The solid (red) line represents the theoretical predictions obtained in Sec.~\ref{sec:perturbative}. The vertical dash-dotted line indicates the boundary for phase separation at the mean-field level, Eq.~\eqref{condition_MF_eq_masses}, while the vertical dashed line refers to the beyond-mean-field stability condition \eqref{condition_full_equal}. Inset: total energy per fermion. Error bars are smaller than symbol size.}
  \label{fig:BMF12B}
  \end{figure}

\begin{figure}[tbp]
    \centering
    \includegraphics[width=\columnwidth]{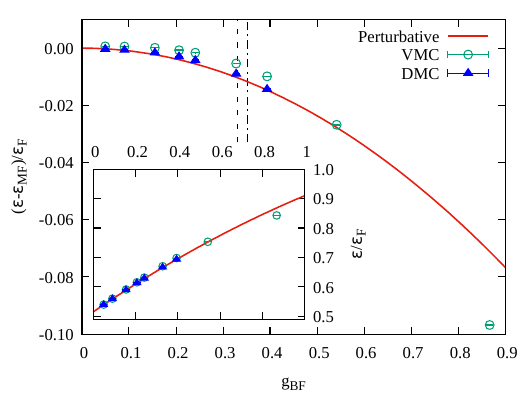}
    \caption{Same as in Fig.~\ref{fig:BMF12B}, but for  bosonic concentration $x_2 \simeq 0.490$
and bosonic repulsion $g_{\rm BB} \simeq 6.2  \cdot 10^{-2}$ .}\label{fig:BMF24B}
\end{figure}

\subsection{Results}\label{sec:results}
In this work, we focus on two reference bosonic concentrations $x_1=12/49 \simeq 0.245$ and $x_2=24/49 \simeq 0.490$, which represent mixtures with small and high bosonic densities, respectively.
The results for the effective masses are shown in Fig.~\ref{fig:EffectiveMass}. They are obtained from the fitting scheme described in Sec.~\ref{subsec:finitesize}, using numbers of fermions $N_{\rm F}=21,29,37,45,49,61,69,81$ (which correspond to closed shells for 2D, as discussed in Sec.~\ref{sec:qmc}) and numbers of bosons chosen so as to get concentrations close to the two reference concentrations $x_1$ and $x_2$, respectively. In the low interaction regime, the QMC results for both concentrations barely differ from the noninteracting value $m^*=m$. For increasing interaction, although the QMC results qualitatively follow Eq.~\eqref{eq:perturbativemass}, some discrepancy between the perturbative predictions and the QMC points is manifest. This might be due to a not sufficiently accurate nodal surface, which could be improved by introducing backflow correlations. For $g_{\rm BF}\gtrsim 0.35$ the disagreement is even more evident. This could be related both to the deficiency of perturbation theory in the strongly interacting regime, and to the limits of our extrapolation scheme for strong BF interactions, where, presumably, fermions can no longer be described with Fermi liquid theory. In fact, in such regime phase separation is expected, and the homogeneous Fermi liquid is at best only a metastable state.

We now discuss the zero-temperature equation of state for the two considered bosonic concentrations. When comparing the perturbative prediction \eqref{eq:EnergyPerturbative} for the ground state energy with our QMC results, we find it convenient to consider the energy per fermion 
$\varepsilon = E/N_{\rm F} = {\cal E}/n_{\rm F}$.
In addition, to better visualize the contribution of the term proportional to $g_{\rm BF}^2$ in the perturbative expansion, we write
$\varepsilon =  \varepsilon_{\rm MF}+\Delta \varepsilon$
where the  ``mean-field'' term
\begin{equation}\label{eq:EnergyMF}
    \varepsilon_{\rm MF}\equiv \varepsilon_{\rm F}\left(1/2+g_{\rm BF}x\right)
    +{\cal E}_{\rm B}/n_{\rm F} \;,
\end{equation}
includes the non-interacting ground-state energy of the Fermi component, its mean-field correction due to interaction with bosons, as well the ground-state energy of the boson component in the absence of interaction with fermions (all divided by $N_{\rm F}$).  With this definition, our perturbative expression \eqref{balanced}  yields for the term  $\Delta\varepsilon =\varepsilon-\varepsilon_{\rm MF} $ the expression
\begin{align}\label{eq:second_order}
    \Delta\varepsilon= -\varepsilon_{\rm F} x \, g_{\rm BF}^2 \ln \frac{2 }{ \sqrt{e}}\;
\end{align}
to order $g^2$.

Our QMC results for the beyond mean-field correction $\varepsilon-\varepsilon_{\rm MF}$, in units of the Fermi energy $\varepsilon_{\rm F}$, are shown in Figs.~\ref{fig:BMF12B}-\ref{fig:BMF24B}, together with the perturbative prediction \eqref{eq:second_order}. The shown VMC and DMC results are obtained by applying, for each $g_{\rm BF}$, the finite-size error correction described in Sec.~\ref{subsec:finitesize} to simulations with $N_{\rm F}=49$. This fermion number is chosen because its finite-size correction entails one of the smallest kinetic energy biases $\Delta t(n_{\rm F},N_{\rm F}=49)$. Furthermore, the DMC energies are the result of proper time-step and walker population analyses. The QMC results are in agreement with the analytic perturbative predictions for small BF interaction values. In particular, for the smaller bosonic concentration $x_1$, DMC results agree with the perturbative expansion for $g_{\rm BF} \lesssim 0.3$  while in the case of $x_2$ agreement is found only  for $g_{\rm BF} \lesssim 0.2$.  We also indicate the perturbative stability conditions of the mixture \eqref{condition_MF_eq_masses} with vertical lines: dot-dashed for the mean-field condition of Eq.~\eqref{condition_MF_eq_masses}, and dashed for the second-order condition of Eq.~\eqref{condition_full_equal}. For stronger interactions, at the perturbative level the mixtures lose their homogeneity. Consequently, one would expect that this regime can no longer be efficiently simulated with a translationally invariant and isotropic Jastrow-Slater wavefunction, such the one that we employ. This is the reason why we do not perform DMC simulations for $g_{\rm BF}>0.4$. Notice, however, that the VMC trial wavefunction still yields stable results, even after this coupling value, possibly describing a metastable state adiabatically connected to the uniform phase.

\begin{figure}[tb]
    \centering
    \includegraphics[width=\columnwidth]{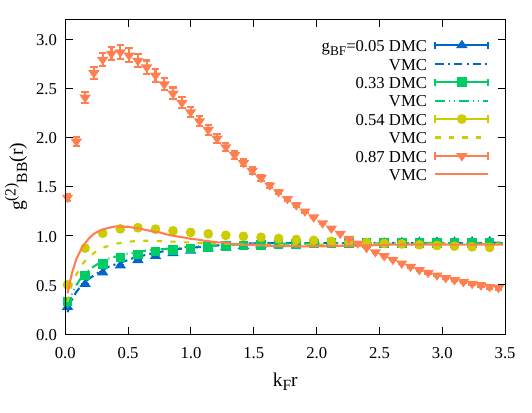}
    \caption{BB pair distribution function for the bosonic concentration $x_1 = 12/49$, with varying BF coupling parameter $g_{\rm BF}$, and bosonic repulsion $g_{\rm BB} \simeq 5.9  \cdot 10^{-2}$, as a function of the dimensionless distance $k_{\rm F}r$. Symbols: DMC results. Lines: VMC results.}\label{fig:gBB12B}
\end{figure}

\begin{figure}[tb]
    \centering
    \includegraphics[width=\columnwidth]{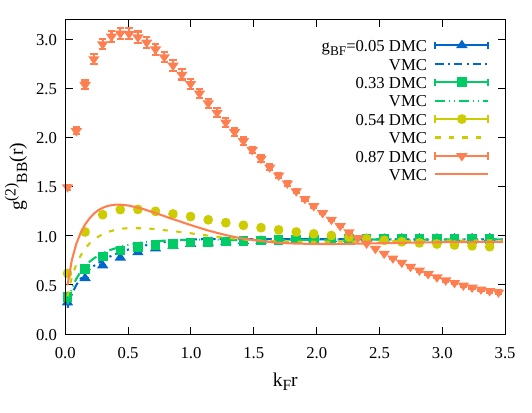}
    \caption{Same as in Fig.~\ref{fig:gBB12B}, but for the bosonic concentration $x_2 = 24/49$ and bosonic repulsion $g_{\rm BB} \simeq 6.2  \cdot 10^{-2}$.}     \label{fig:gBB24B}
\end{figure}

The instability predicted by the perturbative results was indeed experienced during our DMC simulations, as we now describe by performing a qualitative analysis of the BB pair distribution function $g_{\rm BB}^{(2)}(r)$. The results are shown in Figs.~\ref{fig:gBB12B}-\ref{fig:gBB24B}, for two different bosonic concentrations. VMC and DMC results are compared with each other, for different values of the BF coupling.

Since the DMC estimator for pair distribution functions is not \emph{pure} but \emph{mixed}, meaning that it is affected by the used trial wavefunction, usually an extrapolation from the VMC and DMC results is performed, which is valid when their differences are small, or the forward walking technique is used~\cite{Sarsa_QuadraticdiffusionMonte_2002}. Here, instead, we use the discrepancies between the DMC and VMC results as an indication of how suited the employed trial wavefunction is to describe the true groundstate. For both concentrations, and small BF interactions, the BB repulsion suppresses the probability of finding two bosons close to each other. By increasing the distance, the probability density increases until it reaches a plateau value, corresponding to a homogeneous system. This behavior is, in fact, observed for $g_{\rm BF}\lesssim 0.35$. For stronger BF repulsion, the shape of the BB pair distribution function starts to change drastically, presenting a peak for distances less than $1/k_{\rm F}$ and a probability density that decreases for larger distances. Indeed, if the BF repulsion is strong enough, the presence of an effective attraction between the bosons is expected, thus leading to the formation of bosonic clusters. This behavior is much more evident in the DMC results, where the system has evolved towards the true ground state containing the bosonic clusters. Here, the very large discrepancy between the VMC and the mixed DMC estimators points to a state that is significantly different from the employed Jastrow wavefunction, which is translationally invariant and isotropic, since it includes only factors that depend on relative distances. In this inhomogeneous regime, one must be aware that finite-size effects might be significant. The main difference between the two considered boson concentrations is in the prominence of the cluster peaks, which highlights that the higher the concentration the stronger the effect.

\begin{figure}[tb]
    \centering
    \includegraphics[width=\columnwidth]{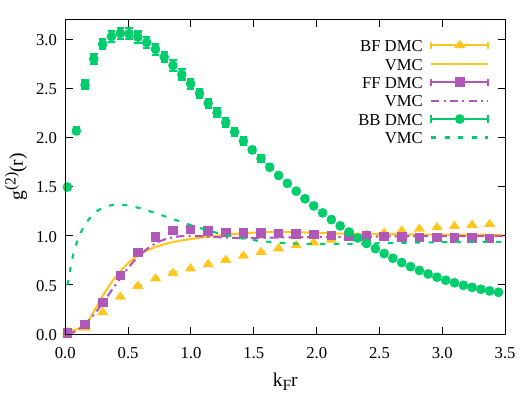}
    \caption{BB, FF and BF pair distribution functions as a function of distance, for $N_{\rm F}=49$, $N_{\rm B}=24$ particles, and interaction parameters $g_{\rm BF} \simeq 8.7 \cdot 10^{-1}$, $g_{\rm BB} \simeq 6.2  \cdot 10^{-2}$. Symbols: DMC results. Lines: VMC results.}\label{fig:g224B}
\end{figure}

To further analyze the clustering behavior of the system, in Fig.~\ref{fig:g224B} we also report the FF and BF pair distribution functions, respectively $g_{\rm FF}^{(2)}(r)$ and $g_{\rm BF}^{(2)}(r)$, for the strong BF repulsion $g_{\rm BF} \simeq 0.87$, by simulating $N_{\rm F}=49$ fermions and $N_{\rm B}=24$ bosons, corresponding to concentration $x_2$. The BB pair distribution function shows typical clustering phenomenology, as discussed above. $g_{\rm FF}^{(2)}(r)$ is instead very similar to the non-interacting case, which is expected, since fermions are the majority species in this mixture and their compressibility is limited by Fermi pressure. Still, the mixed DMC estimator of $g_{\rm FF}^{(2)}(r)$ manifests slightly enhanced oscillations compared with the VMC one. Another convincing hint of clustering comes from the shape of the mixed DMC estimator of $g_{\rm BF}^{(2)}(r)$. A shift in probability towards greater distances between bosons and fermions can be observed, which is further evidence of a reduced uniformity of these two components.
All these considerations qualitatively confirm the presence of phase separation for strong BF repulsion. 

\section{Conclusions}\label{sec:conclusions}

In this work we filled a gap in the theory of 2D dilute BF mixtures by deriving the leading beyond-mean-field contributions to the equation of state, as well as to the fermionic quasi-particle residue and effective mass. For equal masses, and in the repulsive case we have performed QMC simulations, extending to BF mixtures a procedure to correct finite-size effects which was previously used for Fermi-Fermi systems. Our QMC results validate the beyond-mean-field perturbative expansion for the equation of state up to moderate BF repulsion, depending on the bosonic concentration.

The perturbative expansion for the quasi-particle residue and effective mass contains a non-analytic term in the BF coupling (proportional to $|g_{\rm BF}|^{3/2}$) that is absent in the corresponding expansion for a two-component Fermi system \cite{Bloom-1975,Engelbrecht-1992} and which originates from the presence of a condensate in the BF mixture. For the effective mass, we have attempted comparison with the effective mass extracted from  our procedure to correct finite-size effects of QMC simulations, finding in this case only qualitative agreement. Future work will be devoted to increasing the accuracy of the effective mass from QMC simulations of Bose-Fermi mixtures. 

We also investigated the onset of phase separation in repulsive BF mixtures, as predicted by perturbative theory and demonstrated by clustering in the nonperturbative bosonic pair distribution function. While the two results are in qualitative agreement, further work is needed, aiming at quantitatively characterizing the transition from the nonperturbative energetic point of view. This will require a systematic evaluation of the QMC equation of state for different bosonic concentrations in the uniform phase, in order to evaluate a fully nonperturbative stability condition.

Another important research direction will be the verification of the universality of the QMC results, which are here based on a single soft-disk potential, whose effective range could become relevant in the strongly repulsive regime. In particular, while with our model potential we have $n_{\rm F} R_{\rm BF}^2\lesssim 10^{-3}$ before the onset of phase separation, deep in the phase-separated state results can depend on the chosen repulsive model. This is also a regime where in a real atomic system effective repulsion is accompanied by the possible presence of bound states of the full interatomic potential, so that the competition between the phase-separated state and the molecular instability has to be investigated (a problem analogous to the stability of itinerant ferromagnetism in Fermi mixtures \cite{Pilati2010}). Nevertheless, the good agreement between the perturbative and DMC results for the equation of state up to the predicted perturbative phase-separation coupling hints at a negligible role of nonuniversal effects in the uniform phase.

Data for reproducing the figures are available online \cite{DatasetZenodo}.

\begin{acknowledgments}
We acknowledge the use of computational resources from the parallel computing cluster of the Open Physics Hub at the Physics and Astronomy Department in Bologna.
We acknowledge the CINECA award IscrC-BF2D (2021), for the availability of high performance computing resources and support. P.P.~acknowledges financial support from the Italian Ministry of University and Research under project PRIN2022 No.~2022523NA7 and from the European Union - NextGenerationEU through the Italian Ministry of University and Research under PNRR - M4C2 - I1.4 Project CN00000013.
\end{acknowledgments}
\appendix
\section{Quantum scattering theory in two dimensions: main equations}
\label{app:scattering}

We report here the main equations from quantum scattering theory in two dimensions since the 2D case is less standard than the 3D one and notations and definitions are scattered in the literature (and sometimes differ from author to author).

The scattering amplitude $f({\bf k},{\bf k}')$ from the incoming wavevector ${\bf k}$ to the outgoing wavevector ${\bf k}'$ is defined from the asymptotic behavior of the stationary scattering wave function $\psi^+_{\bf k}({\bf r})$ at large distance $r$ from the potential center:
\begin{equation}
\psi^+_{\bf k}({\bf r}) \underset{r\to\infty}{=} \exp(i {\bf k}\cdot {\bf r}) + \frac{f({\bf k},{\bf k}')}{\sqrt{r}}\exp[i(kr+\pi/4)],
\label{scat-amplitude}
\end{equation}
where ${\bf k}'= k {\bf r}/r$ for elastic scattering (see e.g.~\cite{Averbuch-1986}).  With this definition, the differential cross section, like in 3D, is given by
\begin{equation}
\sigma(k,\theta)= |f({\bf k},{\bf k}')|^2,
\end{equation}
 where $\theta$ is the angle between ${\bf k}$ and ${\bf k}'$ 
 The 2D partial-wave expansion reads  
\begin{eqnarray}
f({\bf k},{\bf k}')&=& \sqrt{\frac{2}{\pi k}}\sum_{l=-\infty}^{+\infty} e^{i \delta_l(k)}\sin \delta_l(k) e^{i l \theta}\\
&=& \sqrt{\frac{2}{\pi k}}\sum_{l=-\infty}^{+\infty}\frac{e^{i l \theta}}{\cot \delta_l(k)-i} \, .
\end{eqnarray}
At low energies $\cot\delta_l(k) \propto k^{-2 l}$ for  $l\neq 0$, while (see, e.g., \cite{Hammer_Causalityeffectiverange_2010,Averbuch-1986})
\begin{equation}
\cot\delta_0(k)\underset{k\to 0}{=}\frac{2}{\pi}\ln(k a) + O(k^2).
\end{equation}
We then see that, like in 3D, the s-wave scattering dominates at low energies, so that
\begin{equation}
f({\bf k}',{\bf k}) \underset{k\to 0}{=}\sqrt{\frac{\pi}{2 k}}\frac{1}{\ln (k a) -i \pi/2} .
\end{equation}
The scattering amplitude can be connected to the on-shell two-body $T$-matrix  $t({\bf k}',{\bf k})$, defined by the equation 
\begin{equation}
 t({\bf k}',{\bf k})= T^{\rm 2B}({\bf k}',{\bf k}; z=\epsilon_{\bf k}+i\eta).
 \end{equation}
Here, the (off-shell) two-body $T$-matrix $T^{\rm 2B}({\bf k}',{\bf k}; z)$ is defined as the solution of the Lippmann-Schwinger equation
 \begin{equation} 
 T^{\rm 2B}({\bf k}',{\bf k}; z)=V({\bf k}' -{\bf k} )+ \int\frac{d^2p}{(2\pi)^2}\frac{V({\bf k}'-{\bf p})}{z-\epsilon_{{\bf p}}}T^{\rm 2B}({\bf p},{\bf k}; z)
 \label{eq:c} 
 \end{equation}
 where $\epsilon_{{\bf p}}=\hbar^2 p^2/(2m_r)$, $m_r$ is the reduced mass, $z$ is in general a complex energy, and  $V({\bf q})=\int d^2x e^{-i {\bf q}\cdot {\bf x}}V({\bf x})$ is the Fourier transform of the scattering potential $V({\bf x})$. 
In two dimensions, the connection between $f({\bf k}',{\bf k})$  and $t({\bf k}',{\bf k})$ is given by the relation \cite{Averbuch-1986}
\begin{equation}
   f({\bf k},{\bf k}') = -\frac{m_r}{\hbar^2 \sqrt{2\pi k}}  t({\bf k}',{\bf k}) \,.
\end{equation}
Note that, while for the scattering amplitude $f({\bf k},{\bf k}')$ as defined by Eq.~\eqref{scat-amplitude} one has $|{\bf k}| = |{\bf k}'|$, for the $t$-matrix $t({\bf k}',{\bf k})$ one has in general $|{\bf k}| \neq|{\bf k}'|$.

When both momenta tend to zero, one has \cite{Schick-1971,Bloom-1975}
 \begin{equation}
t({\bf k}',{\bf k})\underset{k, k'\to 0}{=}-\frac{\pi\hbar^2}{m_r}\frac{1}{\ln (k a) -i \pi/2}
\label{t2D} \,,
 \end{equation}
which is the fundamental equation to construct the perturbative expansion of the many-body T-matrix for a dilute BF mixture.

 \section{Perturbative expansion of the many-body T-matrix}
 \label{app-T-matrix}
In this appendix we derive an expansion for the many-body $T$-matrix $\Gamma$ describing the interaction between bosons and fermions to second order in the 2D gas parameter $g_{\rm BF}$.
 For convenience, we set $\hbar = 1$. 

The relevant momenta in our dilute system are of the order of the Fermi momentum $k_{\rm F}$, which is small in comparison with the momentum scale $1/R$ set by the range of the BF interaction. We can thus
use the low-momenta expression (\ref{t2D}) for the on-shell two-body $t$-matrix. Moreover, by introducing the dimensionless momentum variable $\kappa= k/k_{\rm F}$ and using the perturbative assumption $|g_{\rm BF}| \ll 1$, one has
\begin{eqnarray}
    t(\textbf{k}',\textbf{k}) &=& -\frac{\pi}{m_r}\frac{1}{\ln (\kappa k_{\rm F} a_{\rm BF}) -i\pi/2} \\
    &\simeq& \frac{\pi}{m_r} g_{\rm BF}\left[ 1+g_{\rm BF}(\ln \kappa -i\pi/2) \right] 
\end{eqnarray}
where we have used that $\ln(k_{\rm F}a_{\rm BF})$ is the dominant term in the denominator (for both attractive and repulsive cases).

Recalling Eq.~(\ref{gamma-fin}) for the many-body $T$-matrix $\Gamma$, 
and expanding it to second order in $g_{\rm BF}$, one obtains
\begin{widetext}
   \begin{equation} \label{eq:T2DPoleUp}
\Gamma(k;\bar{P})=\frac{\pi}{m_r} g_{\rm BF}\left[1+g_{\rm BF}\left(\ln \kappa -i\frac{\pi}{2}\right) + \left( \frac{\pi}{m_r} g_{\rm BF}\right)\int\!\!\! \frac{ d\textbf{p}}{(2\pi)^2}  \left(  \frac{ \Theta(|\textbf{P}/2-\textbf{p}|-k_{\rm F})}{P_0 - \frac{\left(\textbf{P}/2-\textbf{p}\right)^2}{2m_{\rm F}} -\frac{\left(\textbf{P}/2+\textbf{p}\right)^2}{2m_{\rm B}} +i\eta}
-\frac{1}{{\textbf{k}}^2/2m_r-\textbf{p}^2/2m_r+i\eta} \right)\right]
\end{equation} 
\end{widetext}
where $\mu_{\rm B}$ has been set to zero within the integral on the left-hand side of Eq.~\eqref{eq:T2DPoleUp}, consistently  with the order $g^2$ of our calculations. We observe that, since the  two-body $t$-matrix at low energy depends only on the (magnitude) of the incoming momentum ${\bf k}$, the many-body $T$-matrix does not depend on the outgoing relative momentum $\mathbf{k}'$.
A change of variable $\mathbf{P/2}-\mathbf{p}\rightarrow - \mathbf{p}$ followed by a transformation to polar coordinates yields
\begin{equation} \label{eq:TMatrix2DCylindrical}
\begin{split}
&\Gamma(k;\bar{P})
=\frac{\pi g_{\rm BF}}{m_r}\Biggr[1+g_{\rm BF}\left(\ln \kappa -i\frac{\pi}{2} +  \int_0^{\infty}\!\!\! d\tilde{p} \frac{\tilde{p}}{\tilde{p}^2-\kappa^2-i\eta} \right.\\
&- \int_1^{\infty} \!\!\! d\tilde{p} \int_0^{\pi} \frac{d\theta}{\pi} \frac{(w+1)\tilde{p}}{\tilde{p}^2(w+1)+\tilde{P}^2-\tilde{P}_0w-i\eta+ 2\tilde{P}\tilde{p} \cos\theta} \Biggr)\Biggr].
\end{split}
\end{equation}
where we have introduced the dimensionless variables $\tilde{p}=p/k_{\rm F}$, $\Tilde{P}=P/k_{\rm F}$, $\nu=\omega/\varepsilon_{\rm F}$ and $\tilde{P}_0=P_0/\varepsilon_{\rm F}$.

Integration over the angle yields
\begin{equation} \label{eq:IntCosTheta}
\int_0^{\pi} \frac{d\theta}{\pi}\frac{1}{z+C\cos \theta}=\frac{{\rm sgn}({\rm Re} z)}{\sqrt{z^2-C^2}},
\end{equation}
with ${\rm Im} z \neq 0$ and $C$ real. In the present case
\begin{equation}
{\rm Re} z=\tilde{p}^2(w+1)+\tilde{P}^2-\tilde{P}_0w.
\end{equation}
One thus gets
\begin{eqnarray}
&&\Gamma(k;\bar{P})
=\frac{\pi g_{\rm BF}}{m_r}\Biggr\{1+g_{\rm BF}\Biggr[ \ln \kappa -i\frac{\pi}{2} \nonumber\\ 
&&+ \int_0^{\infty}\!\!\!\!\! d\tilde{p} \Biggr(\frac{\tilde{p}}{\tilde{p}^2-\kappa^2-i\eta} -\frac{\Theta(\tilde{p}-1){\rm sgn}(\tilde{p}^2a + c'))a \tilde{p}}{\sqrt{a^2 \tilde{p}^4+2b\tilde{p}^2+c^2}} \Biggr)\Biggr]\Biggr\} ,\phantom{aaaa}
\label{eq:TMatrix2Dafterangle}
\end{eqnarray}
where
\begin{equation} \label{eq:newvariables}
\begin{split}
&a = w+1\\
&b = \tilde{P}^2 (w-1) - \tilde{P}_0 (w^2+w) - i\eta \equiv b' - i\eta\\
&c =   \tilde{P}^2 - \tilde{P}_0 w - i\eta \equiv c' -i\eta.
\end{split}
\end{equation}
Using
\begin{eqnarray}
\int \!\! \frac{\tilde{p}}{\tilde{p}^2-\kappa^2-i\eta}d\tilde{p} &=&\frac{1}{2}\ln(\tilde{p}^2-\kappa^2-i\eta)\\
\int \!\!   \frac{a \tilde{p}}{\sqrt{a^2 \tilde{p}^4+2b\tilde{p}^2+c^2}} d\tilde{p}&=&\frac{1}{2}\ln \left(\tilde{p}^2+\frac{b}{a^2}\right.\nonumber\\
&+&\left.\sqrt{\tilde{p}^4+\frac{2b\tilde{p}^2+c^2}{a^2}} \right) \label{integral2}
\end{eqnarray}
one notices that the dependence of $\Gamma(k;\bar{P}) $ on the relative momentum $k$ disappears, then yielding $\Gamma(k;\bar{P}) = \Gamma(\bar{P})$ with
\begin{equation}
 \Gamma(\bar{P})=\frac{\pi g_{\rm BF}}{m_r}\left[1+\frac{g_{\rm BF}}{2}F_{\Gamma}(\bar{P})\right]
 \label{Gamma-exp}
\end{equation}
where
\begin{eqnarray}
&&F_{\Gamma}(\bar{P})= -\ln 2 + \ln \left(1+\frac{b'}{a^2}-i\eta\right.\nonumber\\
&&+\left.\sqrt{1+\frac{2b'+c'^2}{a^2}-i\eta}\, \right) \phantom{aaa} {\rm if} \; \frac{\tilde{P}_0 w - \tilde{P}^2}{w+1} < 1 
\label{FG1}
\end{eqnarray}
while
\begin{eqnarray}
&&F_{\Gamma}(\bar{P})= -\ln 2 - \ln \left(1+\frac{b}{a^2}+\sqrt{1+\frac{2b+c^2}{a^2}}\right)\nonumber\\
&&+2\ln \left( \tilde{p}_s^2+\frac{b}{a^2}+\sqrt{\tilde{p}_s^4+\frac{2b\tilde{p}_s^2+c^2}{a^2}} \right)\ \phantom{a} {\rm if} \; \frac{\tilde{P}_0 w - \tilde{P}^2}{w+1} > 1 \phantom{aaaaa}
\label{FG2}
\end{eqnarray}
and we have defined  $\tilde{p}_s^2=\frac{\tilde{P}_0 w - \tilde{P}^2}{w+1}$. 
 \section{Perturbative expansion of the fermionic self-energy}
\label{app-selfF}
The two Feynman diagrams of Fig.~\ref{fig:SelfF} yield two different contributions to the self-energy:
\begin{equation} 
\Sigma_{\rm F}(\bar{k})=\Sigma^{({\rm I})}_{\rm F}(\bar{k}) +\Sigma^{({\rm II})}_{\rm F}(\bar{k})
\end{equation}
with
\begin{eqnarray}
&&\Sigma^{({\rm I})}_{\rm F}(\bar{k})=n_0 \Gamma(\bar{k})\\
&&\Sigma^{({\rm II})}_{\rm F}(\bar{k})=i n_0\!\int\!\!\! \frac{d\textbf{p}}{(2\pi)^2} \!\!\int\!\!\! \frac{dp_0}{2\pi} G_{\rm B}^0(\bar{p})  G_{\rm F}^0(\bar{k}+\bar{p}) \Gamma(\bar{k}+\bar{p})^2\phantom{aaa}\label{second}
\end{eqnarray}
where we have defined $\bar{k}\equiv(\mathbf{k},\omega)$.
Let us consider the two terms separately. For the first term, Eq. (\ref{Gamma-exp}) for $\Gamma$ yields
\begin{equation}
\Sigma^{({\rm I})}_{\rm F}(\bar{k}) = n_{\rm B} \frac{\pi g_{\rm BF}}{m_r}\left[1+\frac{g_{\rm BF}}{2}F_{\Gamma}(\bar{k})\right]
\end{equation}
where, consistently with the order of the expansion, we have replaced $n_0$ with $n_{\rm B}$.

The function $F_{\Gamma}(\bar{k})$ is determined  by the expressions (\ref{FG1}) or (\ref{FG2})  (with $\bar{P}$ replaced by $\bar{k}$) depending whether the condition 
\begin{equation}
\frac{\nu w - \kappa^2}{w+1} < 1
\label{condition}
\end{equation}
 is verified or not (where for convenience we have introduced the dimensionless variables $\nu = \omega/\varepsilon_{\rm F}$ and $\kappa = k/k_{\rm F}$).
 
 In order to obtain the chemical potential $\mu_{\rm F}$, the fermion effective mass $m^*$ at $k_{\rm F}$, and the quasi-particle weight $Z(k_{\rm F})$, it is sufficient to know the self-energy $\Sigma_{\rm F}(\mathbf{k},\omega)$ close to $k_{\rm F}$ and for frequencies in a neighborhood of the energy shell $\nu=\kappa^2$. In this case, $F_{\Gamma}(\bar{k})$ is  always determined by expression (\ref{FG1}).
 Indeed, exactly on shell ($\nu=\kappa^2$ ) the condition (\ref{condition}) reads
\begin{equation} \label{eq:ConditionSigmaFOnShell}
\kappa^2(1-w)+1+w>0
\end{equation}
which, for $w \leq 1$ is always verified, while for $w > 1$ it is verified for
\begin{equation} \label{eq:CondTMatw>1OnShell}
\kappa^2<\frac{1+w}{w-1},
\end{equation}
thus always including a neighborhood of $\kappa=1$. More generally, even off the energy shell, setting $\nu=\kappa^2+\varepsilon$, the condition (\ref{condition}) reads
\begin{equation} \label{eq:ConditionOffShell}
\kappa^2(1-w)+1+w>\varepsilon w.
\end{equation}
For $w\le 1$, the condition (\ref{eq:ConditionOffShell}) is verified for all $\kappa$ if $\varepsilon<1+\frac{1}{w}$, so it is valid in an extended neighborhood of the energy shell $\nu=\kappa^2$.
For $w>1$, it can be proven that the condition (\ref{eq:ConditionOffShell}) is certainly verified for $\kappa < \sqrt{1+1/w}$ and $\varepsilon < \frac{1}{w}$, implying that
there is always a finite neighborhood of $(\kappa=1,\nu=1)$ where the condition (\ref{eq:ConditionOffShell}) is verified.

By using Eq.~(\ref{FG1}) for $F_{\Gamma}(\bar{k})$ we thus obtain
\begin{equation} 
\begin{split}
&\Sigma^{({\rm I})}_{\rm F}(\bar{k}) =\frac{\pi n_{\rm B} g_{\rm BF}}{m_r} + \frac{\pi n_{\rm B} g_{\rm BF}^2}{2m_r}\Biggl[ \ln \Biggl(\frac{(w+1)^2+A(w+1)-2\kappa^2}{(w+1)^2}\\ 
&- i\eta
+\sqrt{1+ \frac{2A}{w+1}-\frac{4\kappa^2-A^2}{(w+1)^2} -i\eta} \Biggr) -\ln 2 \Biggl]
\end{split}
\end{equation}
where $A=\kappa^2-\nu w$.

For the term $\Sigma^{({\rm II})}_{\rm F}(\bar{k})$ as given by Eq.~(\ref{second}), after integrating 
over the frequency $p_0$, replacing  $n_0$ with $n_{\rm B}$, and  $\Gamma^2$ with $(\pi g_{\rm BF}/m_r)^2$  one gets
\begin{equation}
\Sigma^{({\rm II})}_{\rm F}(\bar{k}) =- n_{\rm B}\left(\frac{\pi g_{\rm BF}}{m_r}\right)^2\int\!\! \frac{d\textbf{p}}{(2\pi)^2}\frac{\Theta(k_{\rm F}-|\mathbf{k}+\mathbf{p}|)}{-\omega+\frac{(\mathbf{k}+\mathbf{p})^2}{2m_{\rm F}}-\frac{\mathbf{p}^2}{2m_{\rm B}}+i\eta} .
\end{equation}
After the shift $\mathbf{k}+\mathbf{p} \to \mathbf{p}$, and introducing dimensionless variables:
\begin{equation}
\begin{split}
&\Sigma^{({\rm II})}_{\rm F}(\bar{k}) =\frac{n_{\rm B} }{2 m_r}  g_{\rm BF}^2  \int_0^1 \!\!d\tilde{p} \int_0^{2\pi}\!\!\! d\theta\\ &\times \frac{(w+1)\tilde{p}}{\nu w-\tilde{p}^2(w-1) + \kappa^2-2\tilde{p}\kappa \cos(\theta)+i\eta}.
\end{split}
\end{equation}
 The integral over $\theta$ is solved by using Eq. (\ref{eq:IntCosTheta}), yielding:
\begin{equation} \label{eq:SigmaFIntegralep}
\Sigma^{({\rm II})}_{\rm F}(\bar{k}) =\frac{n_{\rm B} }{2 m_r}  g_{\rm BF}^2 \int_0^1 d\tilde{p} \frac{\sgn(D) 2\pi(w+1)\tilde{p}}{\sqrt{D^2- \left(2\tilde{p}\kappa\right)^2-\sgn(D)\, i\eta }}.
\end{equation}
with 
\begin{equation} 
D=\nu w-\tilde{p}^2(w-1) + \kappa^2.
\end{equation}
For $w \leq 1$, $D$  is always positive in the range of integration for $\nu >-\kappa^2/w$, a condition that is certainly verified in an extended neighborhood of $\nu=\kappa^2$. For $w>1$, on the energy shell $\nu=\kappa^2$, $D$  is always positive if $\kappa^2 > (w-1)/(1+w)$, which is always verified in a neighborhood of $\kappa=1$. More generally, even off the energy shell, one can verify that there is a neighborhood of $\kappa=1$ and $\nu=1$ where $D$  is always positive within the range of integration. For the calculation of the physical quantities of interest in this work, we can then replace $\sgn(D)$ with $1$ in Eq. (\ref{eq:SigmaFIntegralep}) for all values of $w$ and proceed with the calculation of the integral over $\tilde{p}$. 
By defining $B=\kappa^2+\nu w$, one has
\begin{align} \label{eq:SigmaF2DIntegralp01}
&\Sigma^{({\rm II})}_{\rm F}(\bar{k}) =  g_{\rm BF}^2 \frac{n_{\rm B}(w+1) }{2 m_r}   \int_0^1 d\tilde{p} \frac{2\pi \tilde{p}}{\sqrt{a^2\tilde{p}^4+2b\tilde{p}^2+c}} 
\end{align}
where $a=w-1$, $b=-B(w-1)-2\kappa^2$,  $c=B^2-i\eta$.
 The integral over $\tilde{p}$ is solved by using Eq.~(\ref{integral2}),
yielding 
\begin{eqnarray}
\label{eq:SelfFII2D}
&&\Sigma^{({\rm II})}_{\rm F}(\bar{k})= g_{\rm BF}^2 \frac{\pi n_{\rm B}}{2m_r} \frac{w+1}{w-1} \Biggr\{ \ln \Biggr[B(w-1)-(w-1)^2+2\kappa^2\nonumber\\
&&-(w-1)\sqrt{(w-1-B)^2-4\kappa^2-i\eta}\Biggr] -\ln (2\kappa^2)\Biggr\}.
\end{eqnarray}
By summing  (\ref{eq:SigmaFIntegralep}) and (\ref{eq:SelfFII2D}) one finally obtains Eq.~(\ref{eq:SelfF2DCompleta}) of the main text for $\Sigma_{\rm F}(\bar{k})$.

Finally, we report the expression of the self-energy for the specific case of equal masses $m_{\rm F}=m_{\rm B}=m$, which is obtained by taking the limit $w \to 1$ from the general expression (\ref{eq:SelfF2DCompleta})
\begin{equation} \label{eq:SelfF2Deqmasses}
\begin{split}
\Sigma_{\rm F}(\bar{k})&=\frac{2\pi n_{\rm B} g_{\rm BF}}{m} + \frac{\pi n_{\rm B} g_{\rm BF}^2}{m} 
\Biggl\{ \ln \left(1-\frac{\nu}{2}-i\eta \right.\\
&+\left.\sqrt{1-\nu+(k^2-\nu)^2/4-i\eta}\right) -\ln 2\\ 
&+ \frac{k^2+\nu-\sqrt{(k^2+\nu)^2-4k^2-i\eta}}{k^2}\Biggr\}.
\end{split}
\end{equation}
\vspace{0.3cm}

\section{Trial wavefunction}
\label{app:wavefunction}

For completeness, in this appendix we report details regarding the employed trial or guiding wavefunction. The Jastrow factor has a standard form~\cite{Pilati_QuantumMonteCarlo_2005} that is similar to the starting point of the lowest-order constrained variational (LOCV) approximation~\cite{Yu_Stability_2011}. The two-body correlations $f_{\rm P}(r)$, where ${\rm P}={\rm BF, BB}$, and $r$ is the relative distance between the particles in a ${\rm P}$ pair, are the solutions of the 2D two-body radial Schr\"odinger equation $[-(\hbar^2/2m_{\rm P})r^{-1}\partial_r(r\partial_r)+V_{\rm P}(r)-\epsilon_{\rm P}]f_{\rm P}(r)=0$, with the boundary conditions $f_{\rm P}(\bar{R}_{\rm P})=1$ and $f_{\rm P}^\prime(\bar{R}_{\rm P})=0$. The variational parameters $\bar{R}_{\rm P}$  play the role of healing lengths taking care of the effects of  many-body physics on two-body correlations. The explicit form of the functions is:
\begin{equation}
f_{\rm P}(r) = \begin{cases} 
C_0  I_0(\bar{z}_{\rm P} r ) & r< R_{\rm P}  \\ 
C_1 J_0(z_{\rm P} r) + C_2 Y_0(z_{\rm P} r) & R_{\rm P} \leq r <\bar{R}_{\rm P} \\
1 & r\geq \bar{R}_{\rm P}
\end{cases}
\end{equation}
where $C_0 = [C_1  J_0(z_{\rm P}R_{\rm P})+C_2 Y_0(z_{\rm P}R_{\rm P})]/I_0(\bar{z}_{\rm P}R_{\rm P})$, $C_1 = A \cos{\delta}$, $C_2 = - A \sin{\delta}$, $A = [J_0(z_{\rm P} \bar{R}_{\rm P}) \cos{\delta} - Y_0(z_{\rm P} \bar{R}_{\rm P})\sin{\delta}]^{-1}$, $\tan{\delta}=J_1(z_{\rm P} \bar{R}_{\rm P})/Y_1(z_{\rm P} \bar{R}_{\rm P})$, and $J_n$, $Y_n$, and $I_n$ are, respectively, the order $n$ Bessel functions of first and second kind, and the modified Bessel function of first kind. Continuity and boundary conditions determine the effective two-body energy $\epsilon_{\rm P}$ via $[z_{\rm P} J_1(z_{\rm P}R_{\rm P})+\beta J_0(z_{\rm P}R_{\rm P})]/[z_{\rm P} Y_1(z_{\rm P}R_{\rm P})+\beta Y_0(z_{\rm P}R_{\rm P})] =\tan{\delta}$, where $\beta= \bar{z}_{\rm P} I_1(\bar{z}_{\rm P}R_{\rm P})/I_0(\bar{z}_{\rm P}R_{\rm P})$, $\bar{z}_{\rm P} = \sqrt{\zeta_{\rm P}^2-z_{\rm P}^2}$ and $z_{\rm P}^2=2m_{\rm P} \epsilon_{\rm P}/\hbar^2$. $\zeta_{\rm P}$ is fixed by the scattering length as described in the main text.

\section{Fit of the effective mass}
\label{app:fit}
\begin{figure}[htp]
\centering
\includegraphics[width=\columnwidth]{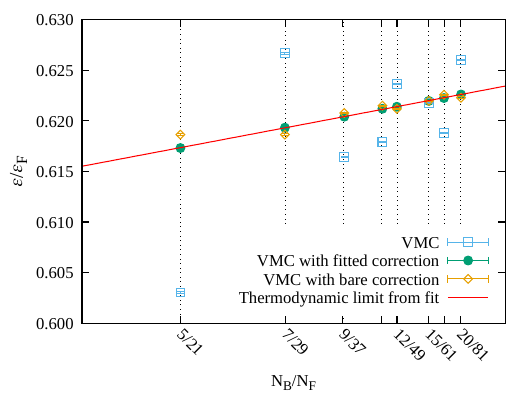}
\caption{Energy per fermion, in units of $\varepsilon_{\rm F}$, as a function of the $N_{\rm B}/N_{\rm F}$ ratio, around the reference bosonic concentration $x_1 \simeq 0.245$, for $g_{\rm BF} \simeq 5.4 \cdot 10^{-1}$. Empty squares: raw VMC results. Filled circles: finite-size corrected VMC results from Eq.~\eqref{eq:TDLimitFitEffectiveMassNB}, employing the fitted inverse effective mass. Empty diamonds: finite-size corrected VMC results from Eq.~\eqref{eq:TDLimitFitEffectiveMassNB}, employing the bare fermionic mass. Solid line: thermodynamic limit of the VMC equation of state in the vicinity of $x_1$, Eq.~\eqref{eq:TDLimitFitEffectiveMassx}.}  
\label{fig:finitesizefit}
\end{figure}
In Fig.~\ref{fig:finitesizefit}, we provide an example of the fitting procedure described in Sec.~\ref{subsec:finitesize}, that aims at estimating the effective masses reported in Fig.~\ref{fig:EffectiveMass}, and thus at providing the main ingredient for the finite-size correction of VMC and DMC results reported in Figures~\ref{fig:BMF12B}-\ref{fig:BMF24B}. 
In particular, we focus on the reference concentration $x_1 \simeq 0.245$ and the coupling $g_{\rm BF} \simeq 5.4 \cdot 10^{-1}$.

We first use the model of Eq.~\eqref{eq:FitEffectiveMass} to fit the VMC results (empty squares) for different $N_{\rm F}$ and $N_{\rm B}$, corresponding to concentrations $x=N_{\rm B}/N_{\rm F}$, with $x \simeq x_1$ - see horizontal axis. The fit yields the parameters $\varepsilon_1$, $c_1$ and $b_1$.
Summing the kinetic energy correction $b_1\Delta t$ to both sides of  Eq.~\eqref{eq:FitEffectiveMass} then brings to two possible estimates of the VMC energy in the thermodynamic limit:
\begin{align}\label{eq:TDLimitFitEffectiveMassNB}
    \varepsilon_{\infty}(n_{\rm F},N_{\rm B}/N_{\rm F})&=\varepsilon(n_{\rm F},N_{\rm F},N_{\rm B})+b_1\Delta t({n_{\rm F},N_{\rm F}})\\
    \bar{\varepsilon}_{\infty}(n_{\rm F},x)&=\frac{\varepsilon_{\rm F}}{2}+\varepsilon_{\rm B}(n_{\rm B})x+\varepsilon_1+c_1(x-x_1)\;.\label{eq:TDLimitFitEffectiveMassx}
\end{align} 

Eq.~\eqref{eq:TDLimitFitEffectiveMassNB} amounts to adding the kinetic energy correction, depending on the fitted inverse effective mass, to each specific VMC result with $N_{\rm F}$ fermions (filled circles in figure). Here, the advantage is that the only needed parameter for the correction is the fitted inverse effective mass $b_1$, while the values of $\Delta t$ are tabulated. We employ this same correction also for DMC simulations with the same particle numbers, having observed that an analogous fitting procedure for DMC (not shown) yields a consistent value for $b_1$. 

Alternatively, Eq.~\eqref{eq:TDLimitFitEffectiveMassx} provides the thermodynamic limit of the VMC equation of state for generic $x$ close to $x_1$ (solid line), which is valuable for considering concentrations not corresponding to specific $N_{\rm B}/N_{\rm F}$ ratios. However, the $\varepsilon_1$, $c_1$ fitted values are only consistent with VMC. Providing a similar expression for DMC would require applying an analogous fitting procedure to DMC simulations, which are much more computationally expensive than VMC calculations.

In the figure, we also show the VMC results with a ``bare'' finite-size correction (empty diamonds), given by Eq.~\eqref{eq:TDLimitFitEffectiveMassNB} in which we set $b_1=1$. It is apparent that employing the fitted $b_1$ yields a much better consistency between the corrected results (filled circles) and the thermodynamic limit equation of state for VMC (solid line).

\end{document}